\documentclass[12pt,letterpaper]{article}

\pdfoutput=1

\usepackage{graphics,epsfig,graphicx,color}
\graphicspath{{/EPSF/}{../Figures/}{Figures/}}
\usepackage{subfigure}

\usepackage{amssymb,latexsym}

\usepackage{setspace}

\usepackage{amsmath}

\usepackage{mathrsfs,amsfonts}

\usepackage{amsthm,amsxtra}

\usepackage{mathtools}

\usepackage[final]{showkeys}

\usepackage{stmaryrd}

\usepackage{algorithm}
\usepackage{algorithmicx}
\usepackage{algpseudocode}

\usepackage[openbib]{currvita}

\newif\ifPDF
\ifx\pdfoutput\undefined
	\PDFfalse
\else
	\ifnum\pdfoutput > 0
		\PDFtrue
	\else
		\PDFfalse
	\fi
\fi

\ifPDF
	\usepackage{pdftricks}
	\begin{psinputs}
		\usepackage{pstricks}
		\usepackage{pstcol}
		\usepackage{pst-plot}
		\usepackage{pst-tree}
		\usepackage{pst-eps}
		\usepackage{multido}
		\usepackage{pst-node}
		\usepackage{pst-eps}
	\end{psinputs}
\else
	\usepackage{pstricks}
\fi

\ifPDF
	\usepackage[debug,pdftex,colorlinks=true, 
	linkcolor=blue, bookmarksopen=false,
	plainpages=false,pdfpagelabels]{hyperref}
\else
	\usepackage[dvips]{hyperref}
\fi

\usepackage{fancyhdr}

\usepackage{comment}

\usepackage[toc,page]{appendix}

\usepackage{cancel}

\usepackage{multirow}

\newcommand{\dsum}{\displaystyle\sum}

\newcommand{\eps}{\varepsilon}

 \newcommand{\bbE}{\mathbb E}

\newcommand{\bbR}{\mathbb R}

\newcommand{\bnu}{{\boldsymbol \nu}}

 \newcommand{\bx}{\mathbf x}

\newcommand{\cA}{\mathcal A} 
  
\newcommand{\cE}{\mathcal E}

\newenvironment{keywords}
{\noindent{\bf Key words.}\small}{\par\vspace{1ex}}
\newenvironment{AMS}
{\noindent{\bf AMS subject classifications 2010.}\small}{\par}

\newcommand{\wt}{\widetilde}

\newcommand{\pt}[1][]{\frac{\partial#1}{\partial t}}

\newcommand{\disp}{\displaystyle}
\newcommand{\tr}{\top}

\renewcommand\arraystretch{1.5}

\makeatletter
\newsavebox{\@brx}
\newcommand{\llangle}[1][]{\savebox{\@brx}{\(\m@th{#1\langle}\)}%
  \mathopen{\copy\@brx\kern-0.5\wd\@brx\usebox{\@brx}}}
\newcommand{\rrangle}[1][]{\savebox{\@brx}{\(\m@th{#1\rangle}\)}%
  \mathclose{\copy\@brx\kern-0.5\wd\@brx\usebox{\@brx}}}
\makeatother

\title{Quantitative Photoacoustic Imaging of Two-photon Absorption}

\author{
	Patrick Bardsley\thanks{
		The Institute for Computational Engineering and Sciences (ICES),
		University of Texas, Austin, TX 78712; 
		bardsley@ices.utexas.edu .
	}
	\and 
	Kui Ren\thanks{
		Department of Mathematics and ICES,
		University of Texas, Austin, TX 78712; 
		ren@math.utexas.edu .
	}
	\and 
	Rongting Zhang\thanks{
		Department of Mathematics, 
		University of Texas, Austin, TX 78712; 
		rzhang@math.utexas.edu .
	}
}
    
\begin{document}

\maketitle



\begin{abstract}
Two-photon absorption photoacoustic tomography (TP-PAT) is a recent hybrid imaging modality that aims at reconstructing two-photon absorption properties of heterogeneous media from measured ultrasound signals generated by the photoacoustic effect. While there have been extensive experimental studies in recent years to show the great promises of TP-PAT, very little has been done on developing computational methods for quantitative image reconstruction in this imaging modality. In this work, we present a mathematical model for quantitative TP-PAT in diffusive media. We implemented a computational strategy for the reconstruction of the optical absorption coefficients, and provide numerical simulations based on synthetic acoustic data to demonstrate the feasibility of quantitative reconstructions in TP-PAT.
\end{abstract}





\section{Introduction}
\label{SEC:Intro}

Photoacoustic imaging (PAT) is a recent biomedical imaging modality that can provide high-resolution images of optical contrast of heterogeneous media such as biological tissues~\cite{Beard-IF11,BuGrSo-NP09,CoLaBe-SPIE09,KuKu-EJAM08,LiWa-PMB09,OrJaTi-AO97,OrKa-HOBD02,PaSc-IP07,Scherzer-Book10,StMoKo-IEEE16,Wang-Book09,Wang-Scholarpedia14,Wang-PIER14}. In a typical PAT experiment, a short pulse of near-infra-red (NIR) photons is sent into the medium to be probed. Photons then propagate inside the medium and a portion of them gets absorbed during the propagation process. The energy absorbed by the medium leads to the heating of the medium, and the heating then forces the medium to expand. The medium cools down when the remaining photons exit, and the cooling process leads to the contraction of the medium. The expansion and contraction of the medium initializes a pressure change inside the medium which then propagates in the form of ultrasound waves. The ultrasound signals on the surface of the medium are then measured. These measurements are used to infer information on the optical properties of the medium~\cite{ArBeBeCoHuLuOgZh-PMB16,CoFrEm-ACS13,CoLaArBe-JBO12,CoArKoBe-AO06,GaTaWaLi-OL15,PoHeZhNa-JBO15,RaNt-MP07,RaViNt-OL07,RiNt-PRE05,ShHaZe-BOE12,TaCoKaAr-IP12,TiXiFaWa-OL15,UpKrPr-JBO17,WaAn-PMB12,WaSuOrAn-PMB12,YaSuJi-PMB10,Zemp-AO10,ZhSe-OE11}.

We study here two-photon photoacoustic tomography (TP-PAT)~\cite{ChPhTh-APL13,LaLeChChSu-OE14,LaBoGrBuBe-OE13,LaVeBoGaKiBuBe-PPUIS14,ReZh-SIAM17,TsSoOmNt-OL14,UrYiYaZh-JBO14,WiYoNoInChChPaSh-Optica14,YaNaTa-PPUIS10,YaNaTa-OE11,YaTa-PPUIS09,YeKo-OE10}, a variant of PAT that is used to image two-photon absorption properties of tissue-like heterogeneous media. Here by ``two-photon absorption" we mean the absorption event where an electron transfers to an excited state after simultaneously absorbing two photons whose total energy exceed the electronic energy band gap~\cite{Mahr-12,RuPe-AOP10,TaPa-Nature79}. Even though it occurs less frequently in normal biological tissues than its single-photon counterpart (i.e. the absorption event where an electron transfers to an excited state after absorbing the energy of a single photon), two-photon absorption is extremely useful in practice. In recent years, various types of materials with strong two-photon absorptions have been proposed and engineered as exogenous contrast agents for different optical imaging modalities~\cite{DeStWe-Science90,So-ELS02,YiLiAl-AO99,ZiWiWe-NB03}. Many such materials can be tuned to be associated with specific molecular signatures. Therefore, they can be used to visualize particular cellular functions and molecular processes inside biological tissues.

There have been extensive experimental investigations on measuring two-photon absorption properties of various materials using TP-PAT~\cite{BaSoKi-JAP82,LaLeChChSu-OE14,LaBoGrBuBe-OE13,YaHaSaMiNiMaHaTaTa-OE14,YaNaTa-PPUIS10,YaNaTa-OE11,YaTa-PPUIS09,YeKo-OE10}. These studies demonstrate the feasibility of TP-PAT in the sense that it is indeed possible to have strong enough photoacoustic effect from two-photon absorption that can be experimentally detected. It has, however, not been satisfactorily demonstrated so far, despite great progress, that one can indeed separate the photoacoustic effect due to single-photon absorption from that due to two-photon absorption to have better quantitative reconstruction of two-photon absorption from measured ultrasound data. In the rest of this paper, we demonstrate, computationally, through a model-based reconstruction algorithm, that it is possible to get quantitative reconstructions of both single-photon and two-photon absorptions and therefore separate them.

\section{The mathematical models}
\label{SEC:Model}

The main physical processes involved in a TP-PAT experiment are the propagation of near infra-red photons and the propagation of ultrasound signals in the underlying medium. In optically heterogeneous media such as the biological tissues, it is well established now that the propagation of NIR photons can be modeled with a diffusion equation for the local density of photons~\cite{BaRe-IP11,CoArKoBe-AO06,ShHaZe-BOE12,Zemp-AO10}. The main difference between TP-PAT and the regular PAT is that two-photon absorption, in addition to single-photon absorption, needs to be considered in the model for light propagation. Let us denote by $\Omega \subseteq\bbR^d$ ($d \ge 2$) the medium to be probed, and denote by $u(\bx)$ the density of photons at position $\bx\in\Omega$. We then have that $u(\bx)$ solves the following nonlinear diffusion equation~\cite{ReZh-SIAM17}:
\begin{equation}\label{EQ:TP Diff}
	\begin{array}{rcll}
  	-\nabla\cdot \gamma(\bx) \nabla u(\bx) + \sigma(\bx) u(\bx) + \mu(\bx)|u|u(\bx) &=& 0, &\mbox{in}\ \ \Omega\\
       u + \kappa \gamma \dfrac{\partial u}{\partial \nu} &=& g(\bx), &\mbox{on}\ \ \partial\Omega
	\end{array}
\end{equation}
where $\nabla$ is the usual gradient operator with respect to the spatial variable $\bx$, and the function $g(\bx)$ models the incoming NIR illumination source on the boundary $\partial\Omega$. The function $\gamma(\bx)$ is the diffusion coefficient of the medium, and the functions $\sigma(\bx)$ and $\mu(\bx)$ are respectively the single-photon absorption and the (intrinsic) two-photon absorption coefficients. The total two-photon absorption strength is given by the product $\mu(\bx) |u|$ where the absolute value operation is taken to ensure that the total two-photon absorption strength is non-negative, a property that needs to be preserved for the nonlinear diffusion model~\eqref{EQ:TP Diff} to correctly reflect the physics. The unit outer normal vector at point $\bx$ on the boundary $\partial\Omega$ is denoted by $\bnu(\bx)$, and the notation $\dfrac{\partial u}{\partial \nu}=\bnu \cdot\nabla u$ is used in the Robin boundary condition. The coupling parameter $\kappa$ in the boundary condition is the rescaled extrapolation length. Its value depends on many parameters, and can be explicitly calculated in specific settings~\cite{DaLi-Book93-6}.

The main difference between the nonlinear diffusion model~\eqref{EQ:TP Diff} and the classical linear diffusion model is the extra term $\mu(\bx)|u|u(\bx)$ that models the two-photon absorption mechanism. This nonlinear term makes the model~\eqref{EQ:TP Diff} harder to solve computationally than the classical linear diffusion model. It is important to emphasize that this nonlinear diffusion model is indeed a well-posed mathematical model that admits a unique solution for a given illumination source $g$ under classical assumptions on regularities of the coefficients and the domain. Classical numerical discretization schemes, such as finite element and finite difference methods, can be used to discretize the equation. Iterative schemes such as the Newton's method can be used to solve the resulting nonlinear algebraic system; see more detailed discussions in~\cite{ReZh-SIAM17}. Let us also mention that simplified versions of this nonlinear diffusion equation have been proposed in previous studies in TP-PAT; see for instance~\cite{ChPhTh-APL13}.

The initial pressure field generated by the photoacoustic effect in TP-PAT is the product of the Gr\"uneisen coefficient of the medium, $\Gamma$, and the total energy absorbed locally by the medium, $\sigma u + \mu |u| u$. Note that here the total absorbed energy consists of two components, the contribution from single-photon absorption, $\sigma u$, and the contribution from two-photon absorption, $\mu |u| u$. Therefore, we write the initial pressure field as~\cite{FiScSc-PRE07,ReZh-SIAM17}:
\begin{equation}\label{EQ:TP Ini Pres}
	H(\bx) = \Gamma(\bx) \Big[\sigma(\bx) u(\bx) + \mu(\bx) |u| u(\bx)\Big], \qquad \bx\in \Omega.
\end{equation}
where the Gr\"uneisen coefficient is non-dimensionalized, and it describes the efficiency of the photoacoustic effect of the underlying medium.

The change of pressure field generates ultrasound waves that propagate following the standard acoustic wave equation, the same model equation for ultrasound propagation in the regular PAT~\cite{FiScSc-PRE07}:
\begin{equation}\label{EQ:TP Wave}
	\begin{array}{rcll}
		\dfrac{1}{c^2(\bx)}\dfrac{\partial^2 p}{\partial t^2} -\Delta p &=& 0, 
			&\text{in}\ (0, +\infty) \times \bbR^d \\
		p(t,\bx)= \chi_\Omega H(\bx),\ \  
		\dfrac{\partial p}{\partial t}(t,\bx) &=& 0,&\text{in}\ \ \{t=0\}\times\bbR^d
	\end{array}
\end{equation}
where $p(t,\bx)$ is the pressure field, and $c(\bx)$ is the speed of the ultrasound waves. In most biological applications of PAT, the ultrasound speed $c$ is assumed known and is often taken as the speed of ultrasound in water since most biological tissues behave like water to ultrasound waves. The function $\chi_\Omega$ is the characteristic function of the domain $\Omega$. It should be understood as the extension operator that extends the initial pressure field inside the medium $\Omega$ to the whole space $\bbR^d$, that is,
\[
	\chi_\Omega H(\bx) = \left\{
	\begin{array}{rl}
		H(\bx), & \bx\in\Omega\\
		0, & \bx \in\bbR^d\backslash\Omega
	\end{array}
	\right. .
\]

The acoustic datum measured in TP-PAT is the ultrasound signal on the surface of the medium, $p_{|(0, T]\times \partial\Omega}$, for time $T$ sufficiently long, and very often, we need to measure data that are generated from multiple illumination sources. From the measured data, we are interested in reconstructing the physical coefficients $(\Gamma, \gamma, \sigma, \mu)$ of the underlying medium. Note that among all the coefficients, the two-photon absorption coefficient $\mu$ is the only new coefficient that appears in TP-PAT. The coefficients $(\Gamma, \gamma, \sigma)$ are also quantities to be reconstructed in the regular PAT~\cite{BaRe-IP11,CoLaBe-SPIE09,CoArKoBe-AO06,LaCoZhBe-AO10,LiWa-PMB09,PuCoArKaTa-IP14,ReGaZh-SIAM13,ReZhZh-IP15,SaTaCoAr-IP13,Wang-PIER14,Zemp-AO10,ZhAn-SPIE06,ZhZhZhGa-IP14}. Mathematical analysis in~\cite{ReZh-SIAM17} shows that one can not simultaneously reconstruct all four coefficients $(\Gamma, \gamma, \sigma, \mu)$ even from data collected from multiple illuminations, if all illumination sources have the same optical wavelength. We will therefore only focus on the two absorption coefficients $(\sigma, \mu)$ in the rest of the paper. The reconstruction of all four coefficients using multispectral data following the ideas in~\cite{BaRe-IP12,CoArBe-JOSA09,CoLaBe-SPIE09,LaCoZhBe-AO10,RaDiViMaPeKoNt-NP09,RaViNt-OL07,ShCoZe-AO11,YuJi-OL09} will be the subject of a future work.

\section{A two-step reconstruction method}
\label{SEC:Two-Step}

We now present a numerical method for the reconstruction of the absorption coefficients $(\sigma, \mu)$. We follow the standard two-step procedure in quantitative PAT image reconstructions. In the first step, the qualitative step, we reconstruct the initial pressure field, $H$ in~\eqref{EQ:TP Ini Pres}, from measured acoustic data using the wave equation~\eqref{EQ:TP Wave}. In the second step, the quantitative step, we reconstruct the absorption coefficients from the initial pressure field $H$ using the nonlinear diffusion equation~\eqref{EQ:TP Diff}. We emphasize that the reason for choosing this two-step reconstruction strategy, instead of the more recent one-step algorithms such as those in~\cite{DiReVa-IP15,PuCoArGoKaTa-IEEE16}, is that the two-step method allows us to avoid solving the nonlinear diffusion equation~\eqref{EQ:TP Diff} in the quantitative reconstruction step in this specific setup; see more discussions in Section~\ref{SUBSEC:Quant}.

\subsection{Qualitative step: reconstructing initial pressure fields}

In the qualitative step of TP-PAT, we aim at reconstructing the initial pressure field $H$ from measured datum $p_{|(0, T]\times \partial\Omega}$. This step is the same as that in the regular PAT, which has been extensively studied in the past. Many efficient algorithms have been proposed; see for instance~\cite{BuMaHaPa-PRE07,CoArBe-IP07,FiHaRa-SIAM07,HaScSc-M2AS05,HaZa-IP10,Hristova-IP09,HuXiMaWa-JBO13,KuKu-EJAM08,Kunyansky-IP08,Nguyen-IPI09,PaNuHaBu-PIS09,QiStUhZh-SIAM11,ScAn-JOSA11,Treeby-JBO13,WaAn-PMB12,ZhSe-OE11} for an incomplete list of works in this direction. We implement here a simple least-square based algorithm for the reconstruction.

To simplify the presentation, let us denote by $\cA$ the linear operator that takes the initial pressure field $H(\bx)$ to the acoustic field on the boundary $\partial\Omega$, i.e.,
\begin{equation}\label{EQ:Data EQ}
	p(t, \bx)_{|(0,T]\times \partial\Omega}=\cA H.
\end{equation}
Our objective is to invert the operator $\cA$ to find $H$ for a given measurement $p^*(t, \bx)_{|(0,T]\times \partial\Omega}$. We solve this problem in the least-square sense, that is, we search for $H$ as the minimizer of the misfit functional
\begin{equation}\label{EQ:Obj Func}
\Phi (H) := \int_0^{T}\int_{\partial\Omega} (p-p^*)^2 d\bx dt \equiv \|\cA H -
p^*\|_{L^2((0,T]\times \partial\Omega)}^2.
\end{equation}
Standard least-square theory then implies that the minimizer $H$ solves the the normal equation
\begin{equation}\label{EQ:Normal}
\cA^\top \cA H = \cA^\top p^*,
\end{equation}
where $\cA^\tr$ denotes the $L^2$-adjoint of the operator $\cA$. We therefore need to invert the self-adjoint operator $\cA^\top \cA$ to find $H$.

We solve the normal equation~\eqref{EQ:Normal} using the Conjugate Gradient method~\cite{Bjorck-Book96}. In a nutshell, the method seeks a solution of the normal equation by iteratively choosing ``conjugate'' or ``$\cA^\tr\cA$-orthogonal'' directions, and minimizing the magnitude of the residual, $\|\cA H-p^*\|_{L^2((0, T]\times\partial\Omega)}^2$, in each of these conjugate directions. 

More precisely, let $H_k$ be the value of $H$ at iteration $k$, and let $\{h_i\}_{i=1}^k$ be the set of ``$\cA^\tr\cA$-orthogonal'' directions constructed in the first $k$ iterations. The directions $\{h_i\}_{i=1}^k$ satisfy the $\cA^\tr\cA$ orthogonality relation, $\forall\ 2\le i\le k$:
\[
	\langle h_i,\cA^\tr\cA h_j \rangle_{L^2(\Omega)}= 0,\ \ \forall\ 1\le j\le i-1.
\]
We now search for an update of $H_k$ in the direction $h_k$ such that the residual is minimized after the update. That is, we minimize the residual $\Psi(\alpha)$ over $\alpha$ with:
\[
\begin{aligned}
	\Psi(\alpha)=\|\cA(H_k+\alpha h_k)-p^*\|_{L^2((0, T]\times\partial\Omega)}^2 &=
\langle
\cA(H_k+\alpha h_k-H),\cA(H_k+\alpha h_k-H)\rangle_{L^2((0, T]\times\partial\Omega)}\\
&= \langle H_k+ \alpha h_k - H,
\cA^\top \cA(H_k+\alpha h_k-H)\rangle_{L^2(\Omega)}
\end{aligned}
\]
where it was recalled $p^* = \cA H$. The optimality condition immediately gives that the step length at iteration $k$ is:
\[
\alpha_k = \frac{\langle
h_k,\cA^\top \cA(H_k-H)\rangle_{L^2(\Omega)}}{\langle
h_k,\cA^\top \cA h_k\rangle_{L^2(\Omega)}} = \frac{\langle
h_k,s_k\rangle_{L^2(\Omega)}}{\langle
h_k,\cA^\top\cA h_k\rangle_{L^2(\Omega)}},\ \ \mbox{with}\ \ s_k=\cA^\top [\cA(H_k-H)].
\]
Note that if we define $r_k=\cA(H_k-H)=\cA H_k-p^*$ as the residual of the original problem at step $k$, then $s_k=\cA^\top r_k$ is simply the so-called normal residual corresponding to $r_k$. 
The updated value $H_{k+1}$ is then obtained as
\[
H_{k+1} = H_{k}+\alpha_k h_k,
\]
while the normal equation residual $s_k$ is updated as
\[
s_{k+1} = s_k - \alpha_k \cA^\top \cA h_k.
\]
The Conjugate Gradient method updates the search direction following
\[
	h_{k+1} = s_{k+1} +
\frac{\|s_{k+1}\|_{L^2(\Omega)}^2}{\|s_{k}\|_{L^2(\Omega)}^2} h_k.
\]
We summarize the Conjugate Gradient method in Algorithm~\ref{ALG:PAT} following the routine in~\cite{Bjorck-Book96}, with an accuracy tolerance parameter $\eps>0$ and the maximal number of iteration $K$.
\begin{algorithm}
\caption{: CG algorithm for qualitative reconstruction}
\label{ALG:PAT}
\begin{algorithmic}[1]
\State Set parameters $\eps$ and $K$; set $k=0$
\State Set initial guess $H = 0$
\State Evaluate the residual $r = p^* - \cA H$ and the normal residual $s = \cA^\top r$
\State Set initial search directions $h = s$
\State Evaluate the size of normal residual $\gamma = \|s\|_{L^2(\Omega)}^2$  
\While{$k \le K$ and $\gamma/\|\cA^\top p^*\|_{L^2(\Omega)}^2>\eps$}
\State $g = \cA h$
\State $\alpha = \gamma/\|g\|_{L^2((0,T]\times\partial\Omega)}^2$
\State $H = H+\alpha h$
\State $r = r-\alpha g, s = \cA^\top r$
\State $\beta = \|s\|_{L^2(\Omega)}^2/\gamma$
\State $\gamma = \|s\|_{L^2(\Omega)}^2$
\State $h = s+\beta h$
\State $k=k+1$
\EndWhile
\end{algorithmic}
\end{algorithm}

It is often the case that a regularization term is added to the misfit functional $\Phi(H)$ in~\eqref{EQ:Obj Func}. In our implementation, we did not include a regularization term in $\Phi(H)$. When it is needed, the two algorithmic parameters $\eps$ and $K$ can both serve as mechanisms to regularize the reconstruction. We did not pursue in this direction in this study, but we understand that tuning regularization can refine some of the reconstruction results that we show in the next section.

Let us mention that even though the operator $\cA$ and its adjoint operator $\cA^\top$ are called in each iteration of the algorithm, these operators are never explicitly formed in the numerical implementation. We only need to know the actions of these operators on given vectors. For instance, to evaluate $\cA h$ for a given $h(\bx)$, we solve the acoustic wave equation~\eqref{EQ:TP Wave} with initial condition $p(0, \bx) = h(\bx)$ and record the solution on the boundary of $\Omega$: $p(t, \bx)_{|(0, T]\times\partial\Omega}$. To evaluate $\cA^\top r$ for a given $r(t,\bx)$, we first solve the following adjoint wave equation:
\begin{equation}\label{EQ:TP Wave Adj}
	\begin{array}{rcll}
		\dfrac{1}{c^2(\bx)}\dfrac{\partial^2 v}{\partial t^2} -\Delta v &=& 0, 
			&\mbox{in}\ (0, T) \times \Omega \\ 
		v(t,\bx)= 0,\ \ \dfrac{\partial v}{\partial \nu}(t,\bx) &=& r,&\mbox{in} \ (0, T)\times\partial\Omega\\
		v(t,\bx)= 0,\ \  
		\dfrac{\partial v}{\partial t}(t,\bx) &=& 0,&\mbox{in}\ \{t=T\}\times\Omega
	\end{array}
\end{equation}
We then take $\cA^\top r=-\dfrac{\partial v}{\partial t}(0,\bx)$. The derivation of~\eqref{EQ:TP Wave Adj} is straightforward, and has been documented previously~\cite{AmBoJuKa-SIAM10,DiReVa-IP15}, so we omit the details here.

\subsection{Quantitative step: reconstructing absorption coefficients}
\label{SUBSEC:Quant}

The second step, the quantitative step, is to reconstruct the optical coefficients from the initial pressure field $H$ recovered in the first step. In recent years, this step was the subject of many computational studies in the case of the regular PAT; see, for instance, ~\cite{AmBoJuKa-SIAM10,ArBeBeCoHuLuOgZh-PMB16,BaRe-IP11,BaRe-CM11,CoLaArBe-JBO12,CoArKoBe-AO06,CoLaBe-SPIE09,LaCoZhBe-AO10,MaRe-CMS14,PuCoArKaTa-IP14,ReGaZh-SIAM13,ReZhZh-IP15,SaTaCoAr-IP13,ShHaZe-BOE12,Zemp-AO10,ZhAn-SPIE06,ZhZhZhGa-IP14} for a partial list of references. 

Our main objective here is to develop an algorithm to reconstruct the absorption coefficients $(\sigma, \mu)$ in TP-PAT to show that we can separate two-photon absorption from single-photon absorption. We assume that both the Gr\"uneisen coefficient $\Gamma$ and the diffusion coefficient $\gamma$ are known already, for instance from a regular PAT reconstruction.

We assume that we have reconstructed initial pressure fields generated from $J\ge 2$ illuminations sources. We denote by $\{H_j=\Gamma \big[\sigma u_j + \mu |u_j| u_j\big]\}_{j=1}^J$ those initial pressure fields, where $u_j$ denotes the solution of the nonlinear diffusion equation with illumination sources $g_j$ $(1\le j\le J)$.

We first reconstruct from the initial pressure fields $\{H_j\}_{j=1}^J$, using the fact that $\Gamma$ and $\gamma$ are known, the quantities:
\begin{equation}\label{EQ:Absorp Energy Rec}
	\sigma u_j + \mu |u_j| u_j = \dfrac{H_j}{\Gamma},\ \ 1\le j\le J.
\end{equation}
This allows us to replace the term $\sigma u_j + \mu |u_j| u_j$ in the nonlinear diffusion equation~\eqref{EQ:TP Diff} for source $j$ to obtain the following linear diffusion equation for $u_j$ $(1\le j\le J)$:
\begin{equation}\label{EQ:TP Diff j Rec}
-\nabla\cdot(\gamma\nabla u_j) = -\frac{H_j}{\Gamma}\quad \text{in}\ \Omega,\qquad 
u_j+\kappa \frac{\partial u_j}{\partial \nu}= g_j\quad \text{on} \ \partial\Omega .
\end{equation}
We can solve this linear elliptic equation to reconstruct $u_j$, again since $\Gamma$ and $\gamma$ are known. Therefore we can reconstruct the quantities
\begin{equation}\label{EQ:Absorp Coeff Rec}
	\sigma + \mu |u_j|=\dfrac{H_j}{\Gamma u_j},\ \ 1\le j\le J.
\end{equation}
Therefore, at each point $\bx\in\Omega$, we have the following system to determine $\sigma$ and $\mu$:
\begin{equation*}
\begin{pmatrix}
1 & |u_1|\\
\vdots & \vdots \\
1 & |u_J|
\end{pmatrix}
\begin{pmatrix}
\sigma\\ \mu
\end{pmatrix}
=
\begin{pmatrix}
\frac{H_1}{\Gamma u_1}\\
\vdots \\
\frac{H_J}{\Gamma u_J}
\end{pmatrix}.
\end{equation*}
We then reconstruct $(\sigma, \mu)$ by solving this small linear system, in least-square sense, at each point $\bx\in\Omega$, to get
\begin{eqnarray}
\nonumber
\begin{pmatrix}
\sigma\\ \mu
\end{pmatrix}
&=&
\left(
\begin{pmatrix}
1 & \cdots & 1\\
|u_1| & \cdots & |u_J|
\end{pmatrix}
\begin{pmatrix}
1 & |u_1|\\
\vdots & \vdots \\
1 & |u_J|
\end{pmatrix}
\right)^{-1}
\begin{pmatrix}
1 & \cdots & 1\\
|u_1| & \cdots & |u_J|
\end{pmatrix}
\begin{pmatrix}
\frac{H_1}{\Gamma u_1}\\
\vdots \\
\frac{H_J}{\Gamma u_J}
\end{pmatrix} 
\\
\label{EQ:Absorp Final}
&=&
\begin{pmatrix}
J & \sum_{j=1}^J|u_j|\\
\sum_{j=1}^J|u_j| & \sum_{j=1}^J|u_j|^2
\end{pmatrix}^{-1}
\begin{pmatrix}
\sum_{j=1}^J\frac{H_j}{\Gamma u_j}\\
\sum_{j=1}^J\frac{H_j|u_j|}{\Gamma u_j}
\end{pmatrix}.
\end{eqnarray}
We have assumed here that the small $2\times 2$ matrix 
\[
\begin{pmatrix}
J & \sum_{j=1}^J|u_j|\\
\sum_{j=1}^J|u_j| & \sum_{j=1}^J|u_j|^2
\end{pmatrix}
\]
in~\eqref{EQ:Absorp Final} is invertible at each point $\bx\in\Omega$. Theoretical analysis in~\cite{ReZh-SIAM17} shows that one can indeed invert this matrix if the illuminations are selected carefully, that is, the illuminations are sufficiently different from each other. In our numerical experiments, we observe that this matrix is invertible for almost all illuminations that we have tried.

We now summarize the quantitative reconstruction step in Algorithm~\ref{ALG:QPAT}. 
\begin{algorithm}
\caption{: Non-iterative algorithm for quantitative reconstruction}
\label{ALG:QPAT}
\begin{algorithmic}[1]
\For{$j\gets 1, J$}
\State Reconstruct the quantities $\sigma u_j + \mu |u_j| u_j$ following ~\eqref{EQ:Absorp Energy Rec}
\EndFor
\For{$j\gets 1, J$}
\State Solve the diffusion equation~\eqref{EQ:TP Diff j Rec} to reconstruct $u_j$
\EndFor
\For{$j\gets 1, J$}
\State Reconstruct the quantities $\sigma + \mu |u_j|$ following ~\eqref{EQ:Absorp Coeff Rec}
\EndFor
\For{each point $\bx\in\Omega$}
\State Evaluate $\omega=\dsum_{j=1}^J|u_j(\bx)|$, $\theta=\dsum_{j=1}^J |u_j(\bx)|^2$, $\xi=\dsum_{j=1}^J\frac{H_j}{\Gamma u_j}$ and $\zeta=\dsum_{j=1}^J\frac{H_j|u_j|}{\Gamma u_j}$
\State Evaluate $(\sigma, \mu)$ using the formula:
$\begin{pmatrix}
\sigma(\bx)\\ \mu(\bx)
\end{pmatrix}
=
\begin{pmatrix}
J & \omega \\
\omega & \theta
\end{pmatrix}^{-1}
\begin{pmatrix}
\xi\\
\zeta
\end{pmatrix}$
\EndFor
\end{algorithmic}
\end{algorithm}

Let us emphasize two important features of the quantitative reconstruction algorithm. First, even though the reconstruction of the coefficients $(\sigma, \mu)$ from initial pressure field $H$ is a nonlinear inverse problem, our reconstruction method is non-iterative. Therefore there is no convergence issues at all. The method is guaranteed to give the correct reconstruction result.  Second, the reconstruction algorithm is computationally cheap. The major computational cost of the reconstruction algorithm is the solution of the $J$ linear diffusion equations in ~\eqref{EQ:TP Diff j Rec}. The cost in dealing with the algebraic calculations in the rest of the algorithm, in ~\eqref{EQ:Absorp Energy Rec},~\eqref{EQ:Absorp Coeff Rec} and ~\eqref{EQ:Absorp Final}, is almost  negligible.

\section{Numerical implementations}
\label{SEC:Implem} 

We now provide some details on the numerical implementation of the two-step algorithm for quantitative TP-PAT image reconstructions. We limit ourselves to two-dimensional simulations. Nothing changes in three-dimensional case besides the increasing of the computational cost. We use the notational convention $\bx=(x, y)$ for the spatial variable.

Since the units of the physical quantities in the nonlinear diffusion model~\eqref{EQ:TP Diff} and the acoustic wave equation~\eqref{EQ:TP Wave} are very different, we first normalize the problems by taking the following convention. We take the spatial domain $\Omega$ to be the unit square $\Omega = [0,1]^2$, and set the ultrasound speed $c=1$. We set the time interval where the measurements are taken as $(0,T]$ with $T=3$. This convention means that if we take the size of $\Omega$ in units of ${\rm cm}^2$, the ultrasound speed at $1.5\times 10^5 {\rm cm}/{\rm s}$, then $T=3$ equals $20$ microseconds. We observed in our numerical simulations (see discussions in the next section) that these choices of $\Omega$, $c$ and $T$ for the wave equation are sufficient to capture all of the physical wave signals generated by the initial conditions $H(\bx)$ we have tested, at least up to the numerical discretization errors. For the diffusion problem, we set $U_0=10^{11}$ as the characteristic photon density involved in the system, and normalize the solution $u$ and the boundary illumination against $U_0$.

The wave equation~\eqref{EQ:TP Wave} is posed on $\bbR^2$, not inside $\Omega$. We therefore have to make a truncation to have a finite domain for the wave simulation. We do this by using the technique of perfectly matched layers (PML)~\cite{Berenger-JCP94,CoTs-Geophysics01}. We surround our physical domain $\Omega$ with a PML region of thickness $0.2$ to have the computational domain $\wt{\Omega} = [-0.2,1.2]^2$. We use a split-field PML scheme (see, e.g.,~\cite{Berenger-JCP94,CoTs-Geophysics01}). This
scheme reduces to the undamped wave equation in the physical domain
$\Omega$, coupled with a damped wave split field scheme in the PML
region $\wt{\Omega}\setminus\Omega$. Ultimately, we end up solving the system of equations, assumed again that $c=1$,
\[
\renewcommand\arraystretch{2}
\left\{\begin{array}{c}
\disp\pt[p_x] + \tau_xp_x = \frac{\partial v_x}{\partial x},\\
\disp\pt[p_y] + \tau_yp_y = \frac{\partial v_y}{\partial y},\\
\disp\pt[v_x] + \tau_xv_x = \frac{\partial p}{\partial x},\\
\disp\pt[v_y] + \tau_yv_y = \frac{\partial p}{\partial y},
\end{array}
\right. 
\]
where $p = p_x+p_y$, $\tau_x(\bx)$ and $\tau_y(\bx)$ are absorptive terms supported only in the PML region $\wt{\Omega}\setminus\Omega$. In our simulations, we use $\tau_x(\bx) = \tau_x(x,y) = \chi_{x>1}(x)\alpha(x-1)^2+\chi_{x<0}(x)\alpha x^2$ with $\alpha$ a given constant. Similarly, $\tau_y(x,y) = \tau_x(y,x)$. Initial and boundary conditions can be transformed into this first-order formulation in a straightforward way. 

We discretize these equations using standard second-order finite differences in space, and first-order finite differences in time on uniform spatial-temporal grids. The spatial grid covering $\wt\Omega$ consists of $141\times141$ spatial points: 
\[
 \{(x_i,y_j): x_i = i/100, y_j = j/100,\ \ -20 \le i, j\le 120\},
\]
while the temporal grid covering $[0, T]=[0, 3]$ consists of $3001$ grid points:
\[
	\{t_k: t_k= k/1000,\ \  0\le k\le 3000\}.
\]
The velocity fields $v_x$ and $v_y$ are solved for at staggered half-time steps, i.e., at times $\{t_{k+1/2}\}$, using values of the split pressure fields $p_x$ and $p_y$ at the usual time steps $\{t_k\}$. The pressure field $p=p_x+p_y$ is then updated using the velocity fields at these staggered times. Hence the scheme is known as a leapfrog scheme, and reduces to standard second-order finite difference time stepping in the physical domain $\Omega$ where $\tau_x=\tau_y=0$.
With our spatial step size $h=1/100$ and our temporal step size $\Delta t = 1/1000$, we clearly satisfy the CFL condition
\[
\frac{(c\Delta t)^2}{h^2} = \frac{10^4}{10^6} = 10^{-2} <1, 
\]
which is necessary for the explicit finite difference scheme we implemented to be stable.

To solve the adjoint wave equation~\eqref{EQ:TP Wave Adj}, we first perform the change of variable $t'=T-t$ to transform the equation into an initial value (instead of final value) problem. We then apply the same type of spatial-temporal discretizations to the new equation.

The nonlinear diffusion equation~\eqref{EQ:TP Diff} and the linear diffusion equations involved in the reconstruction process, mainly in~\eqref{EQ:TP Diff j Rec}, are all discretized using a standard first-order finite element method with about $12000$ elements on a triangular mesh of $\Omega$. The nonlinear algebra system resulting from the discretization of ~\eqref{EQ:TP Diff} is solved with the Newton's method. In the forward simulation, the initial pressure field $H$ that is needed in the acoustic wave equation~\eqref{EQ:TP Wave} is linearly interpolated from the quadrature points of the triangular elements. In the reconstruction process, the initial pressure field $H$ reconstructed in the qualitative step, i.e. the first step, is interpolated back to the quadrature points of the triangular elements as datum for the quantitative reconstruction step. These interpolation processes induce additional noise in the reconstruction process besides the artificial white noise we add to the acoustic data that we discuss below.

To generate synthetic data, we solve the nonlinear diffusion equation~\eqref{EQ:TP Diff} with the true physical coefficients to generate $H$ and then solve the acoustic wave equation~\eqref{EQ:TP Wave} to produce $p(t,\bx)_{|(0, T]\times\partial\Omega}$. To add noise to the synthetic data, we use the following strategy. At each point $\bx\in\partial\Omega$ where the ultrasound signal is measured, we generate an independent Gaussian
``white-noise'' process $w_\bx(t)$, $t\in(0, T]$, that satisfies
\[
\begin{array}{c}
	\bbE(w_\bx(t)) = 0,\ \ \mbox{and}, \ \ \bbE(w_\bx(t)w_\bx(s)) = \delta(t-s).
\end{array}
\]
We then scale the white noise $w_\bx(t)$ according to the power of the signal $p(t,\bx)$, to generate noisy data $\wt p(t, \bx)$ with a specified noise-to-signal (NSR) ratio $\eta$:
\[
	\wt p(t, \bx) = p(t, \bx) + \eta \left(\frac{\int_0^T
p^2(t', \bx) dt'}{\int_0^T
w_\bx^2(t')dt'}\right)^{1/2} w_\bx(t),\ \ t\in(0, T].
\]

In our numerical simulations in the upcoming section, we set the tolerance level $\eps = 10^{-6}$ in Algorithm~\ref{ALG:PAT} and run this algorithm for a maximum number of $K=1000$ iterations. The quantity $\|\cA H_k - p^* \|^2_{L^2((0, T]\times\partial\Omega)}$ is usually guaranteed to decrease monotonically (see, e.g., the discussion in~\cite[Section 7.4]{Bjorck-Book96}). However, it is clear from our description above that our implementation of the operators $\cA$ and $\cA^\top$ constitute only approximate adjoints of one another due to errors associated with the finite difference approximations and PML region in the wave solver. Therefore, we also force Algorithm~\ref{ALG:PAT} to exit if non-monotonic behavior of $\|\cA H_k - p^* \|^2_{L^2((0, T]\times\partial\Omega)}$ is encountered.

\section{Numerical simulations}
\label{SEC:Num}

We now present some numerical simulations to demonstrate the performance of our quantitative reconstruction strategy. Our main objective is to show that the photoacoustic data measured in TP-PAT allow quantitative separation of the single-photon and the two-photon absorption coefficients. In all the simulation results below, we set the Gr\"uneisen coefficient $\Gamma=1$ and the diffusion coefficient $\gamma(\bx)=0.02+0.01\sin(2\pi y)$. Moreover, we use data collected from four different illumination sources. The first source function takes a constant value the top and right sides of the boundary, and is zero everywhere else. That is,
\[
	g_1(\bx) = \left\{
	\begin{matrix}
		1, & \bx\in(0,1)\times\{1\}\cup \{1\}\times (0,1)\\
		0, & \mbox{otherwise}
	\end{matrix}\right.
\]
The second to fourth sources, $g_2$, $g_3$, and $g_4$, are obtained by rotating $g_1$ by $\pi/2$, $\pi$ and $3\pi/2$ respectively along the boundary. Note again that $g_1$ is normalized against $U_0$ already.
\begin{table}[!ht]
\centering
\setlength{\tabcolsep}{1em}
\renewcommand\arraystretch{1}
\begin{tabular}{c|c|c|c}
NSR & Illumination & $\cE_{L^2}(H)$ & $(\cE_{L^2}(\sigma),\cE_{L^2}(\mu))$\\
\hline\hline
\multirow{4}{*}{$\eta=0.00$} & $g_1$ & $4.05\times 10^{-4}$ & \multirow{4}{*}{$(0.46, 3.33)\times 10^{-2}$}\\
			     & $g_2$ & $6.66\times 10^{-4}$ &\\
			     & $g_3$ & $6.22\times 10^{-4}$ &\\
			     & $g_4$ & $4.96\times 10^{-4}$ &\\ \hline
\multirow{4}{*}{$\eta=0.01$} & $g_1$ & $7.13\times 10^{-3}$ & \multirow{4}{*}{$(1.71, 4.08)\times 10^{-2}$}\\
			     & $g_2$ & $7.30\times 10^{-3}$ &\\
			     & $g_3$ & $8.25\times 10^{-3}$ &\\
			     & $g_4$ & $8.00\times 10^{-3}$ &\\ \hline
\multirow{4}{*}{$\eta=0.05$} & $g_1$ & $1.78\times 10^{-2}$ & \multirow{4}{*}{$(3.80, 6.33)\times 10^{-2}$}\\
			     & $g_2$ & $1.80\times 10^{-2}$ &\\
			     & $g_3$ & $1.77\times 10^{-2}$ &\\
			     & $g_4$ & $1.93\times 10^{-2}$ &\\ \hline
\multirow{4}{*}{$\eta=0.10$} & $g_1$ & $2.68\times 10^{-2}$ & \multirow{4}{*}{$(5.15, 8.01)\times 10^{-2}$}\\
			     & $g_2$ & $2.63\times 10^{-2}$ &\\
			     & $g_3$ & $2.48\times 10^{-2}$ &\\
			     & $g_4$ & $2.55\times 10^{-2}$ &\\ \hline
\end{tabular}
\caption{Quality of reconstructions in Experiment I. Shown are relative $L^2$ errors in the reconstructions of various initial pressure fields (third column) and the corresponding absorption coefficients in Figure~\ref{FIG:True Coeff 2Step} (fourth column) from ultrasound data with different noise levels (controlled with the NSR $\eta$).}
\label{TAB:Exp I}
\end{table}

To measure the quality of the reconstructions, we use relative $L^2$ errors. Let $f$ be a quantity to be reconstructed, $f_t$ its true value and $f_r$ the reconstructed value. Then the relative $L^2$ error of the reconstruction, denoted by $\cE_{L^2}(f)$, is the ratio between the size of the error in the reconstruction and the size of the true quantity. That is,
\[
	\cE_{L^2}(f) = \dfrac{\|f_t-f_r\|_{L^2(\Omega)}}{\|f_t\|_{L^2(\Omega)}}.
\] 

\begin{figure}[htbp]
\centering
\includegraphics[width=0.325\textwidth]{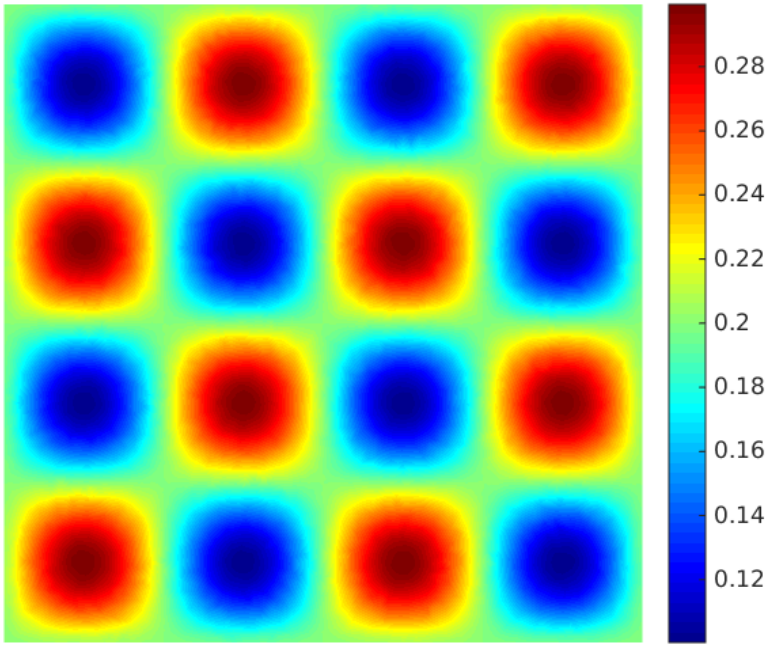} \hskip 1cm
\includegraphics[width=0.325\textwidth]{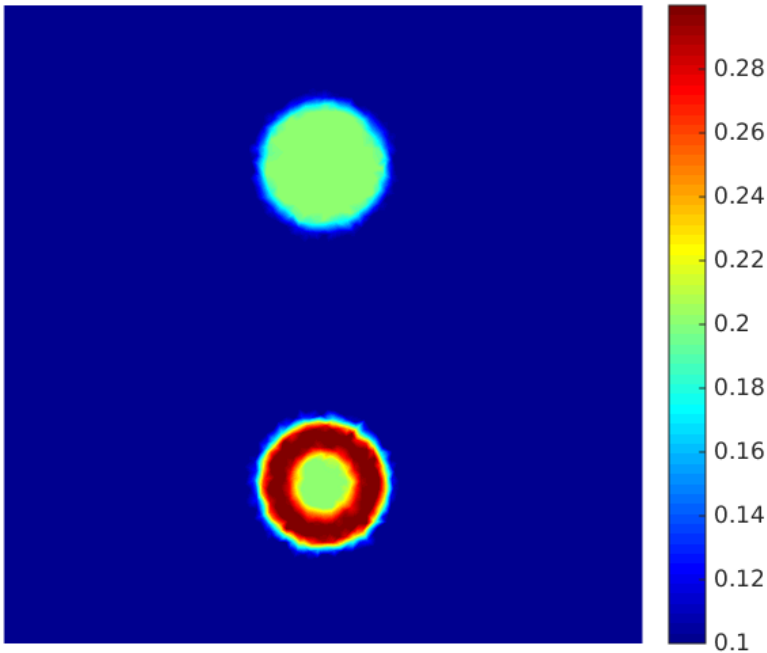}
\caption{The true absorption coefficients, $\sigma$ (middle, $[{\rm cm}^{-1}]$) and $\mu\times U_0$ (right, $[{\rm cm}^{-1}]$), used to generate synthetic data in Experiment I.}
\label{FIG:True Coeff 2Step}
\end{figure}
\begin{figure}[htbp]
\centering
\includegraphics[width=0.325\textwidth]{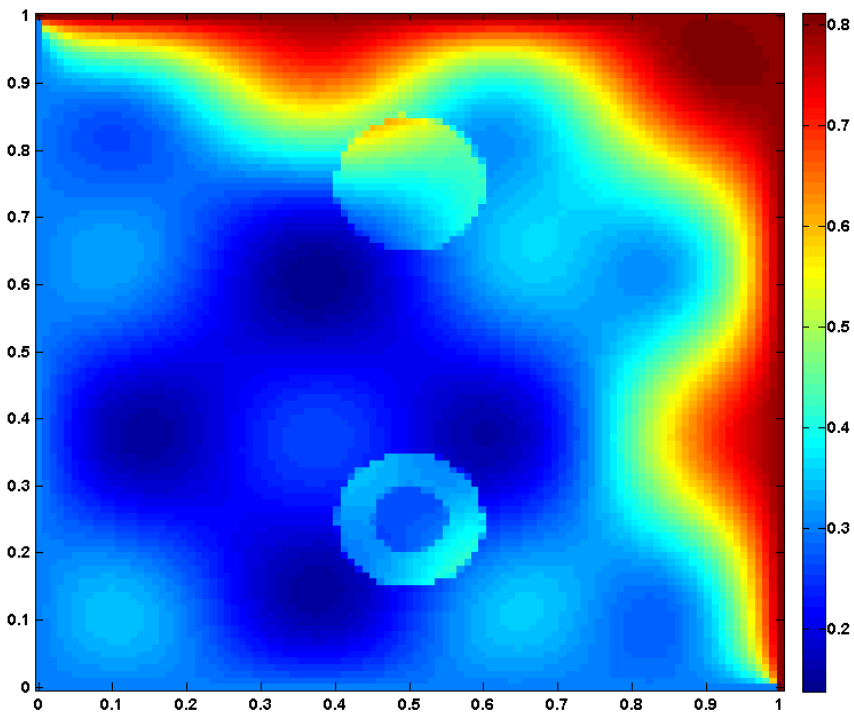}
\includegraphics[width=0.325\textwidth]{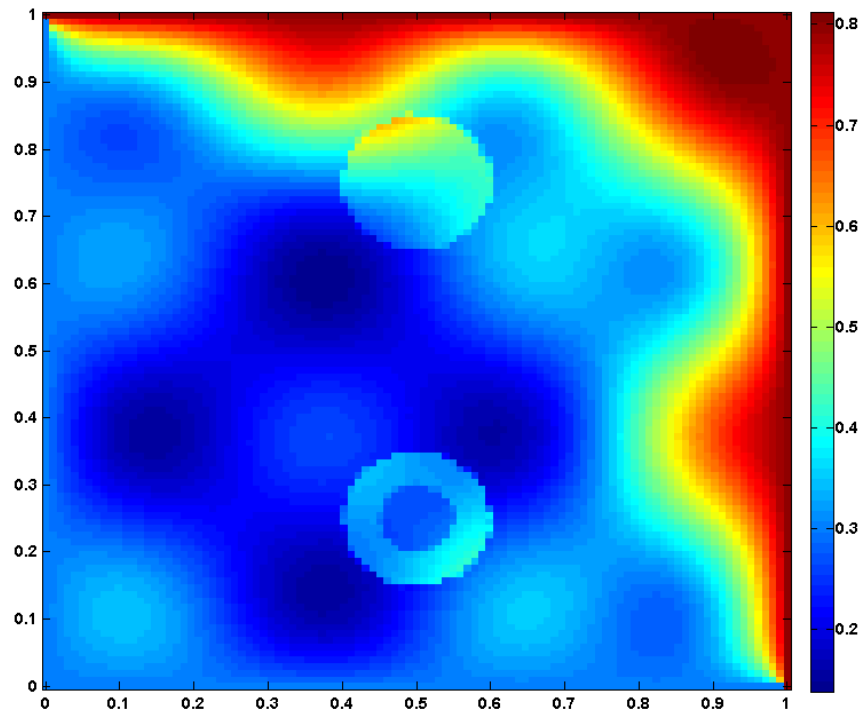}
\includegraphics[width=0.325\textwidth]{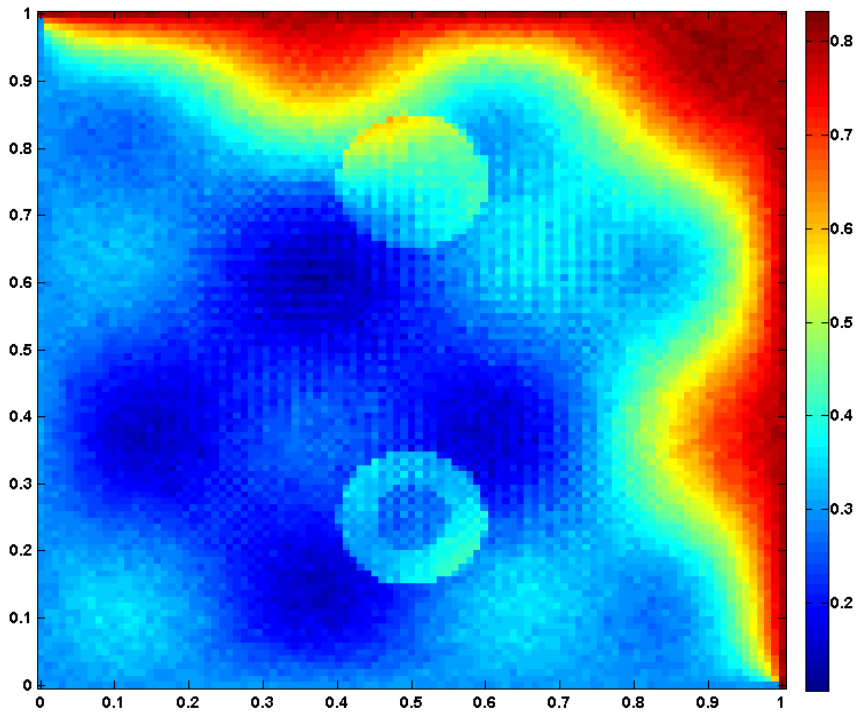}
\caption{The initial pressure field $H(\bx)$ generated from illumination $g_1$ using the true absorption coefficients in Figure~\ref{FIG:True Coeff 2Step} (left), as well as the reconstructions of $H$ using ultrasound data with NSR $\eta=0.0$ (middle, clean data) and NSR $\eta=0.1$ (right).}
\label{FIG:H Recon Illu1}
\end{figure}
\begin{figure}[htbp]
\centering
\includegraphics[width=0.325\textwidth]{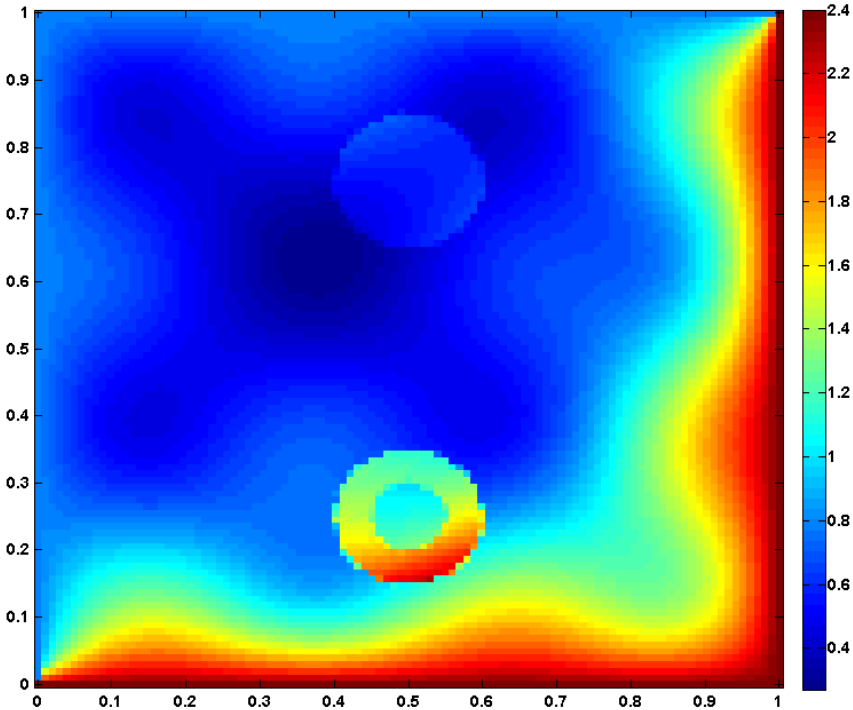}
\includegraphics[width=0.325\textwidth]{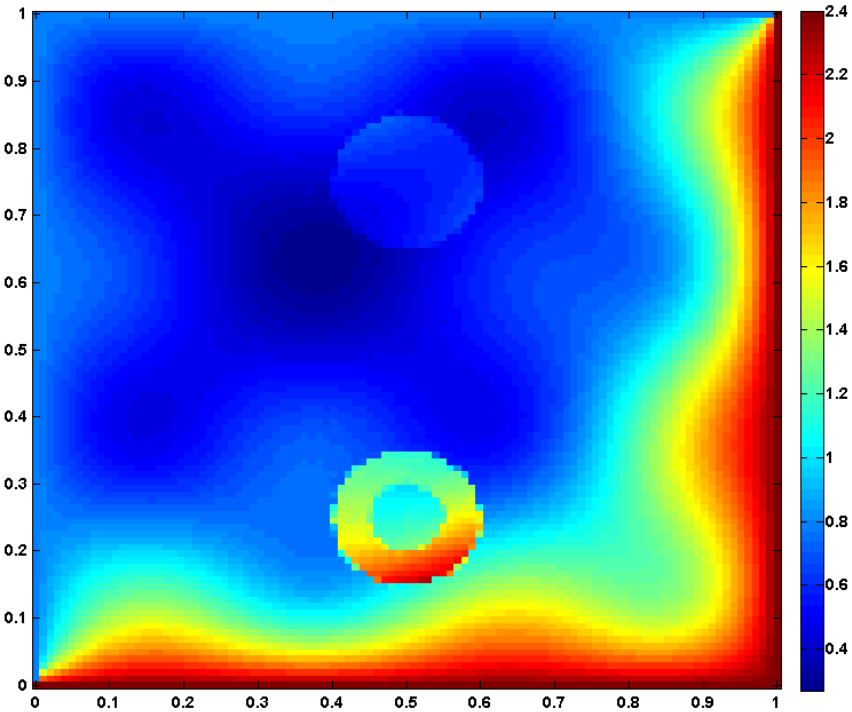}
\includegraphics[width=0.325\textwidth]{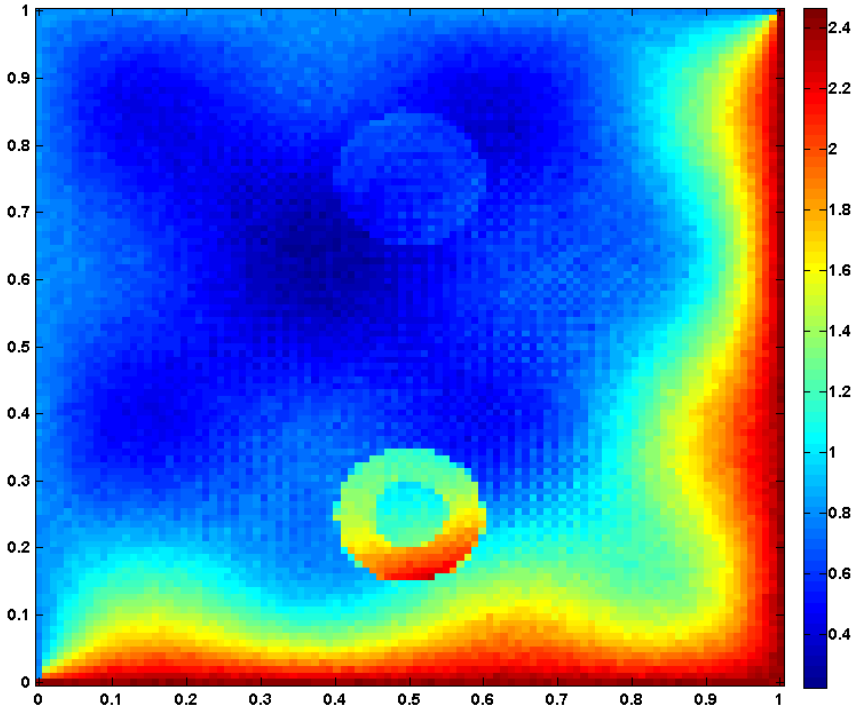}
\caption{The same as Figure~\ref{FIG:H Recon Illu1} except that the illumination source used is $g_2$.}
\label{FIG:H Recon Illu2}
\end{figure}

\paragraph{Experiment I.} In the first group of numerical simulations, we attempt to reconstruct the absorption coefficients $(\sigma, \mu)$ shown in Figure~\ref{FIG:True Coeff 2Step}. We first perform reconstructions, using Algorithm~\ref{ALG:PAT}, on the initial pressure field $H$ generated by the four illumination sources $\{g_i\}_{i=1}^4$ from acoustic data of different noise levels. The quality of the reconstructions, in terms of the relative $L^2$ errors, is summarized in Table~\ref{TAB:Exp I}, third column. We observed, as has been confirmed by many works in the PAT community, the qualitative reconstruction is of high quality. We show in Figure~\ref{FIG:H Recon Illu1} and Figure~\ref{FIG:H Recon Illu2} some reconstructions with illuminations $g_1$ and $g_2$ respectively. Shown are the true initial pressure field $H$, the reconstructed $H$ using clean data (NSR $\eta=0.0$) and noisy data with NSR $\eta=0.1$. The relative $L^2$ errors of the reconstruction in Figure~\ref{FIG:H Recon Illu1} are $0.04\%$ for the case of $\eta=0.0$ and $2.7\%$ for the case of $\eta=0.1$, while these for the reconstructions in Figure~\ref{FIG:H Recon Illu2} are respectively $0.06\%$ for the case of $\eta=0.0$ and $2.6\%$ for the case of $\eta=0.1$. 

Let us mention that even though can see clearly the two-photon absorbing inclusions in the true initial pressure field $H$ and the reconstructed $H$, the true absorption coefficients in Figure~\ref{FIG:True Coeff 2Step} are very different from the $H$ in Figure~\ref{FIG:H Recon Illu1} and Figure~\ref{FIG:H Recon Illu2}. In other words, knowing $H$ does not provide us enough information about the true absorption coefficients unless we perform the next step, the quantitative step, of the reconstruction.

\begin{figure}[htbp]
\centering
\includegraphics[width=0.325\textwidth]{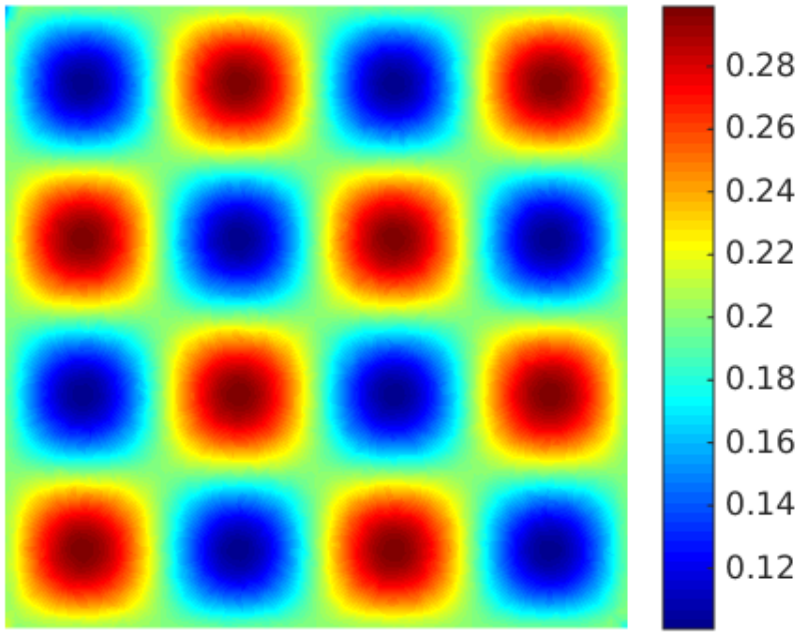}
\includegraphics[width=0.325\textwidth]{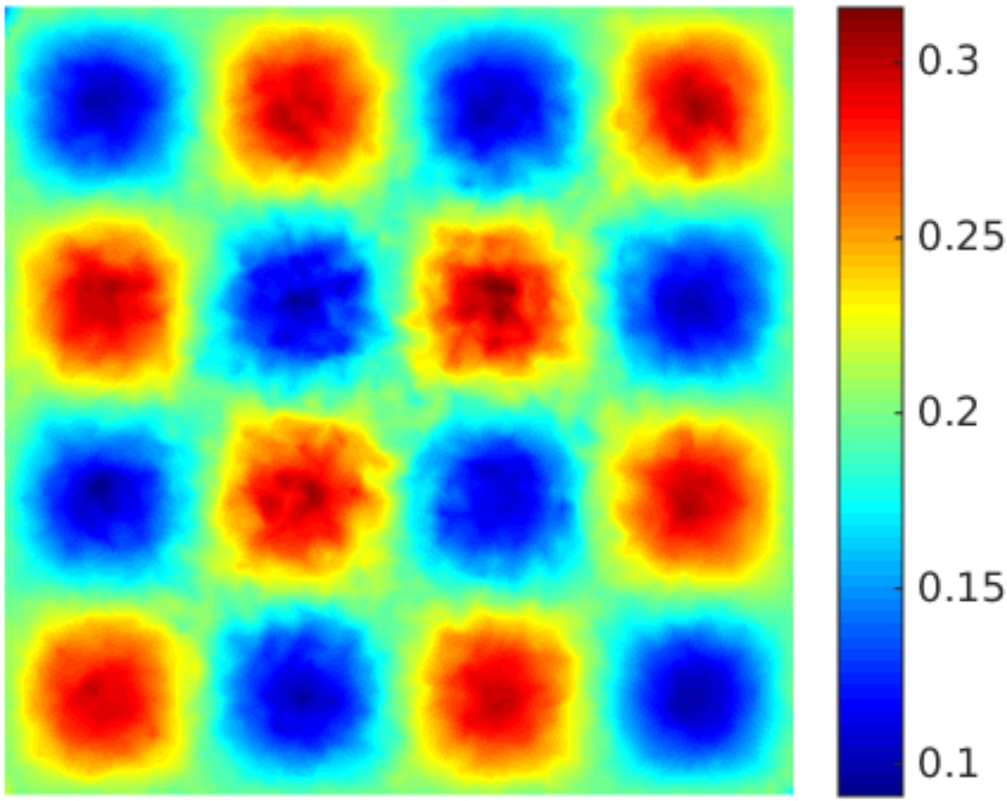}
\includegraphics[width=0.325\textwidth]{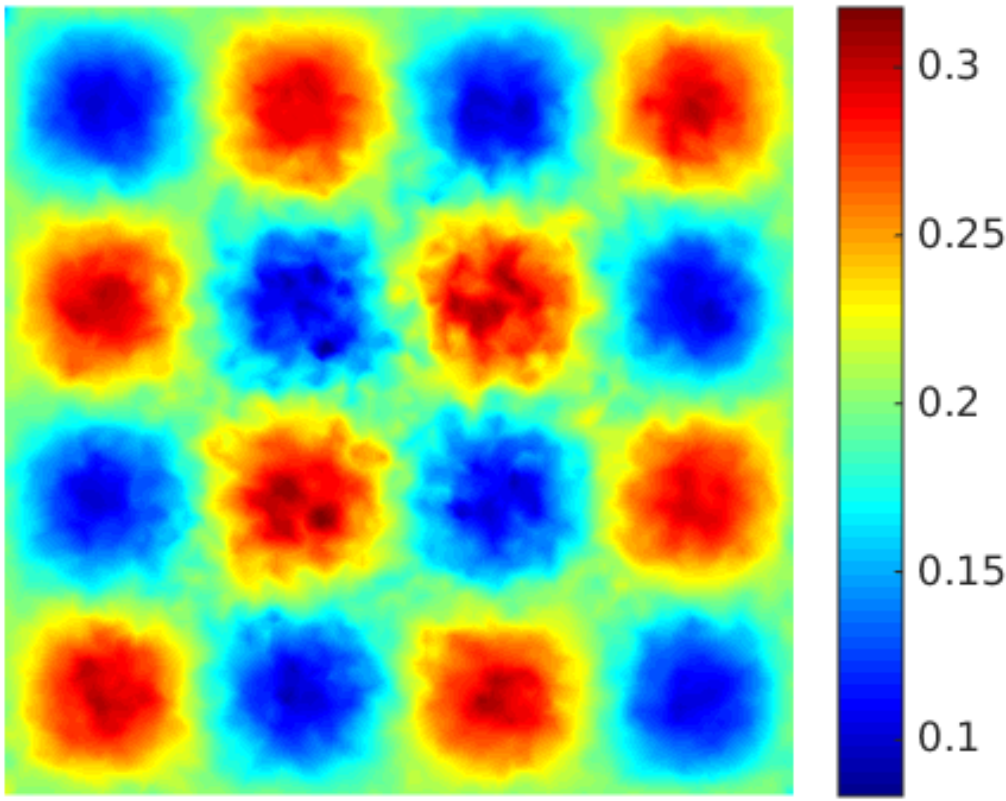}\\
\includegraphics[width=0.325\textwidth]{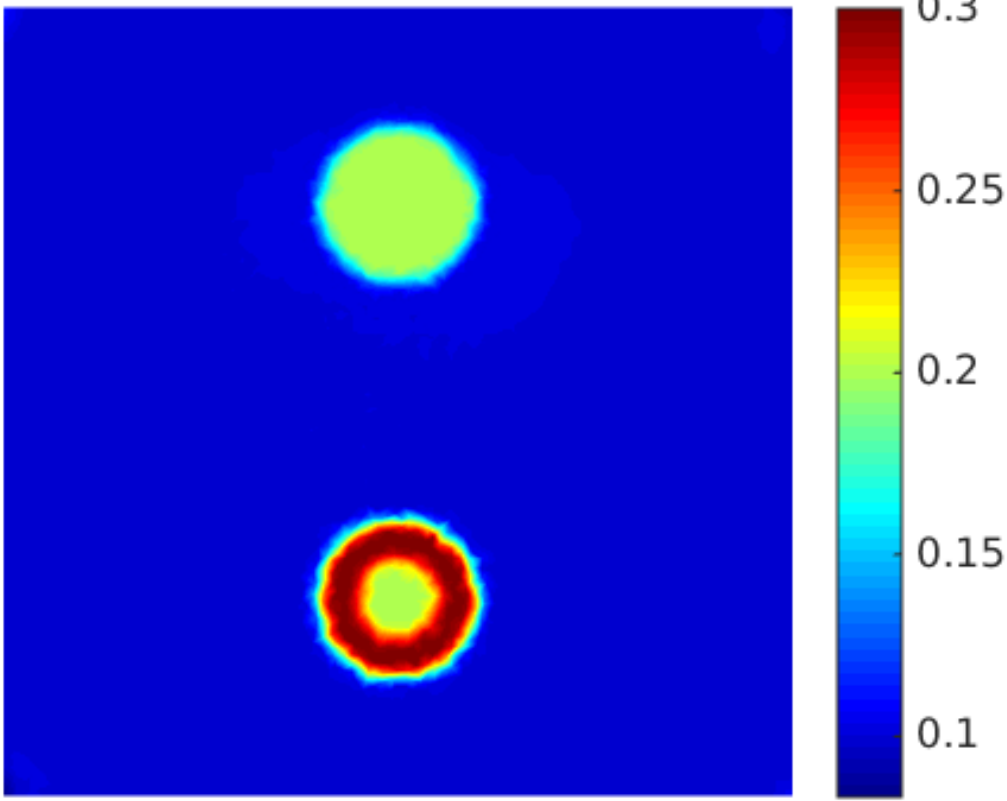}
\includegraphics[width=0.325\textwidth]{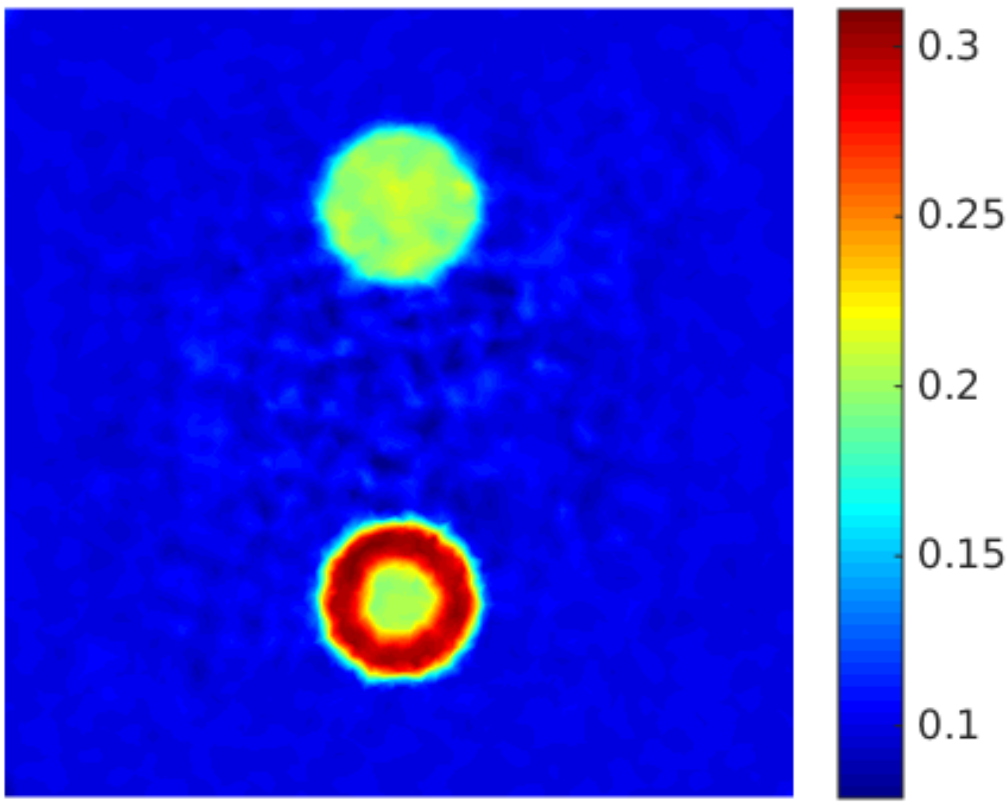}
\includegraphics[width=0.325\textwidth]{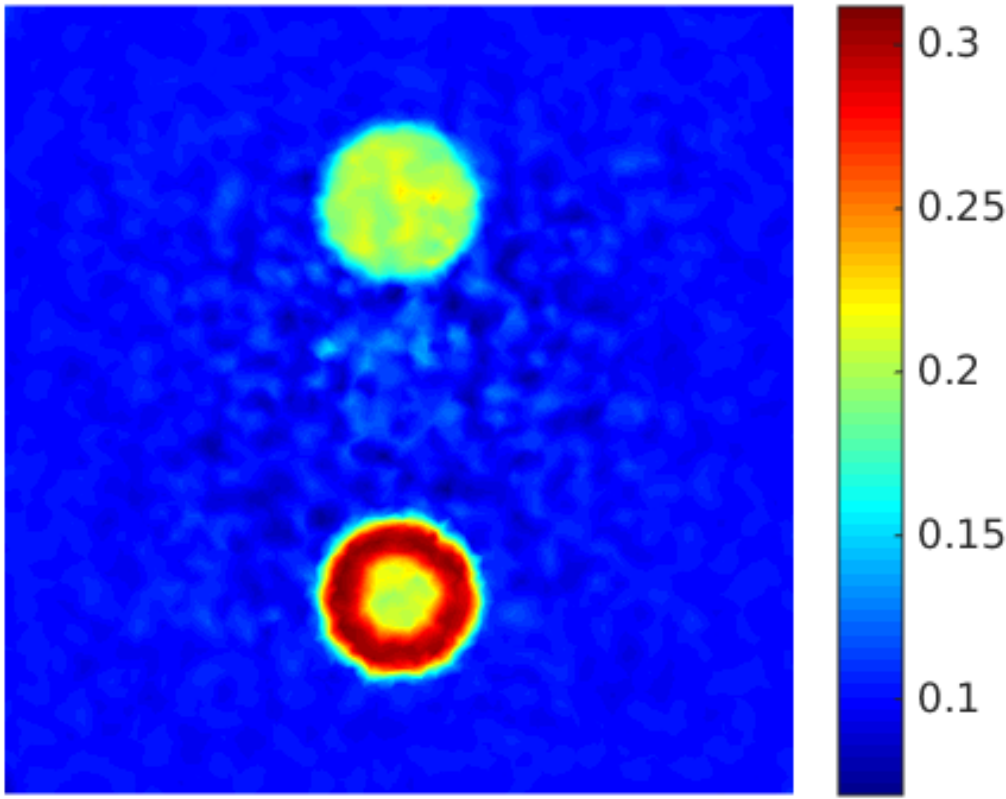}
\caption{The absorption coefficients $\sigma$ (top row) and $\mu\times U_0$ (bottom row) reconstructed using noisy data with NSR $\eta=0.00$, $\eta=0.05$ and $\eta=0.10$ (from left to right).}
\label{FIG:Sigma Mu Direct 2Step}
\end{figure}

In Figure~\ref{FIG:Sigma Mu Direct 2Step}, we show the reconstructions of the coefficient pair $(\sigma, \mu)$ in Figure~\ref{FIG:True Coeff 2Step} from the initial pressure fields we obtained in the qualitative step, using Algorithm~\ref{ALG:QPAT}. Shown, from left to right, are reconstructions using noisy data with NSR $\eta=0.00$, $\eta=0.05$ and $\eta=0.10$ respectively. The quality of the reconstructions is very high with relative $L^2$ errors $(1.71, 4.08)\times 10^{-2}$, $(3.80, 6.33)\times 10^{-2}$, and $(5.15, 8.01)\times 10^{-2}$ respectively; see the fourth column of Table~\ref{TAB:Exp I} for the reconstruction result using data with $\eta=0.01$ which we did not show here since it is too similar to the case with $\eta=0.00$.

The reconstructions in Figure~\ref{FIG:Sigma Mu Direct 2Step} show that by performing quantitative reconstructions, we can separate the two-photon absorption coefficient from the single-photon absorption coefficient from the initial pressure field. This is clearly important for practical applications of TP-PAT where two-photon absorption is the main quantity of interest.

\begin{figure}[htbp]
\centering
\includegraphics[width=0.325\textwidth]{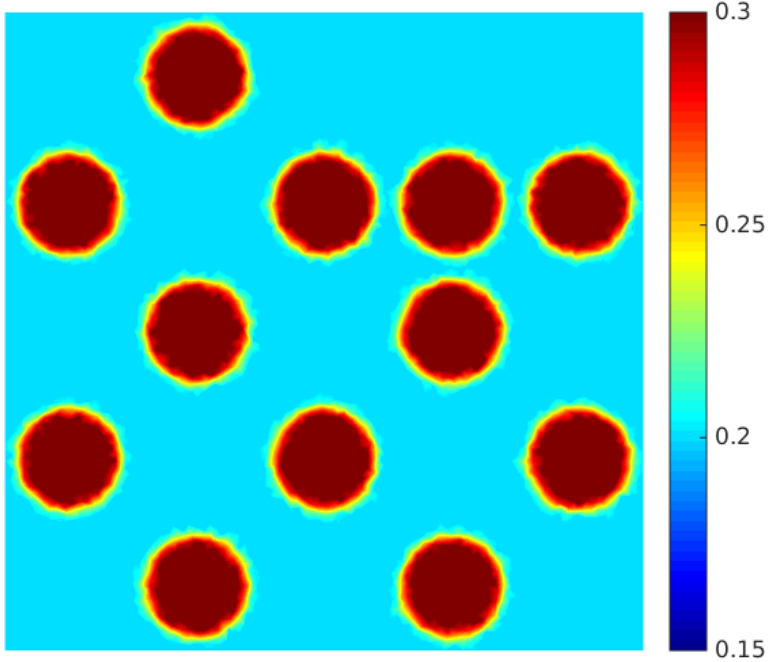}\hskip 1cm
\includegraphics[width=0.325\textwidth]{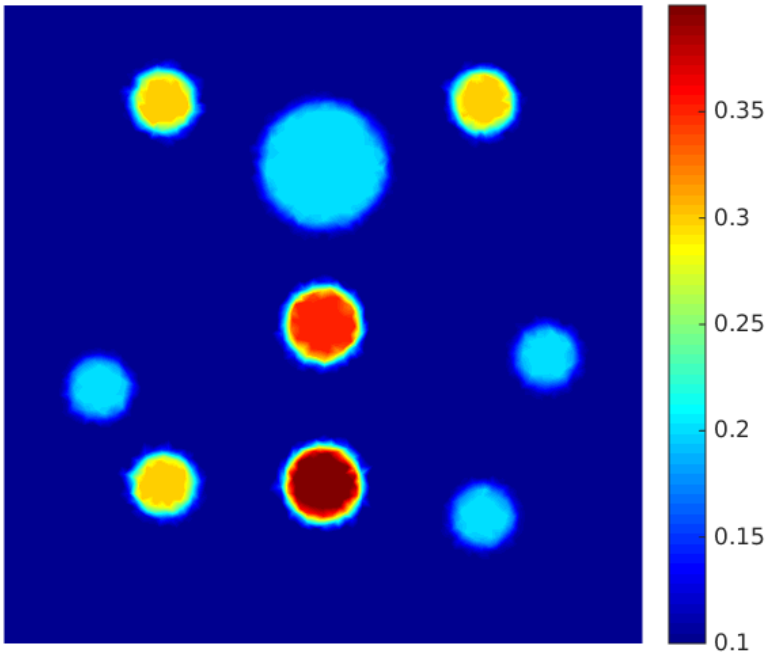}
\caption{The true absorption coefficients, $\sigma$ (middle, $[{\rm cm}^{-1}]$) and $\mu\times U_0$ (right, $[{\rm cm}^{-1}]$), used to generate synthetic data in Experiment II.}
\label{FIG:True Coeff 2Step2}
\end{figure}
\begin{figure}[htbp]
\centering
\includegraphics[width=0.325\textwidth]{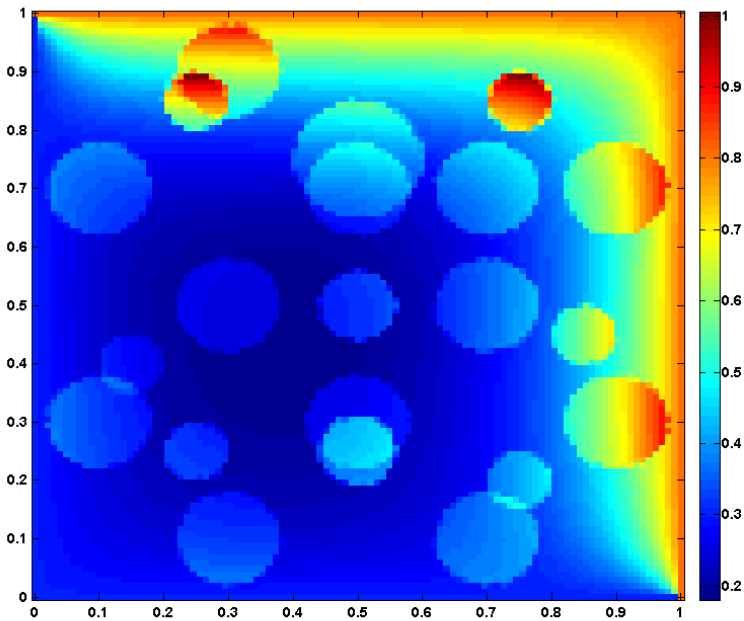}
\includegraphics[width=0.325\textwidth]{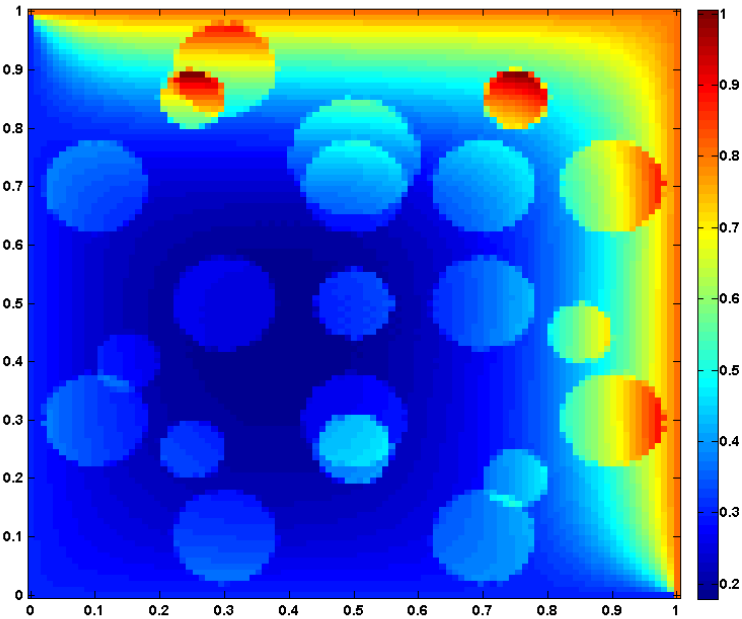}
\includegraphics[width=0.325\textwidth]{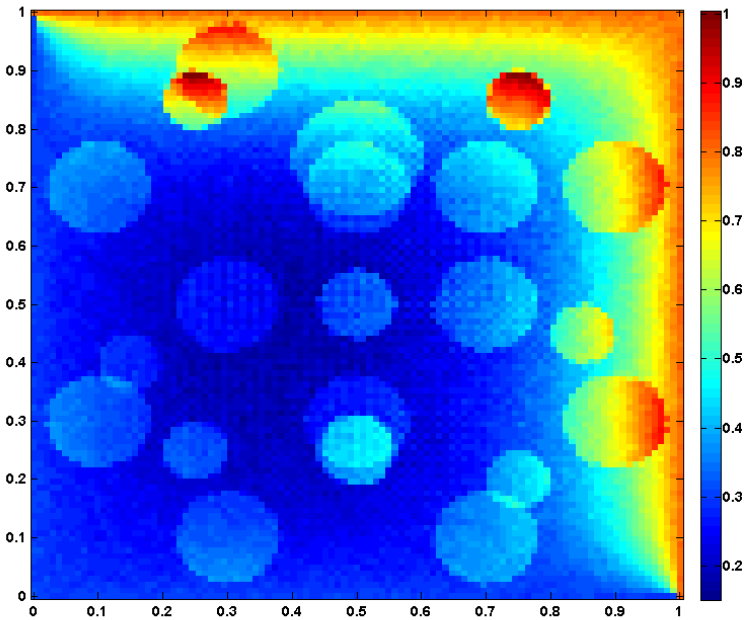}
\caption{The initial pressure field $H(\bx)$ generated from illumination $g_1$ using the true absorption coefficients in Figure~\ref{FIG:True Coeff 2Step2} (left), as well as the reconstructions of $H$ using ultrasound data with NSR $\eta=0.00$ (middle, clean data) and NSR $\eta=0.10$ (right).}
\label{FIG:H Recon Illu5}
\end{figure}

\paragraph{Experiment II.} In the second group of numerical simulations, we study the reconstruction of the absorption  coefficients $(\sigma, \mu)$ shown in Figure~\ref{FIG:True Coeff 2Step2}. In Figure~\ref{FIG:H Recon Illu5}, we present the true initial pressure filed $H$ computed with illumination source $g_1$, and the reconstructions of this $H$ with clean ultrasound data (NSR $\eta=0.00$) and noisy data (NSR $\eta=0.10$). By visual inspection, we can see the presence of both the single-photon absorption and the two-photon absorption inclusions in $H$. The reconstructions are impressively good, with relative $L^2$ errors $9.30\times 10^{-4}$ and $2.51\times 10^{-2}$ respectively for $\eta=0.00$ and $\eta=0.10$, even when $H$ is this complicated. We have also performed similar reconstructions for $H$ generated from the other three illumination sources $g_2$, $g_3$ and $g_4$. The relative errors in the reconstructions are summarized in Table~\ref{TAB:Exp II}, third column. Despite its slight degeneration as noise level increases, the quality of the reconstructions of $H$ remains high at moderate noise levels. 

\begin{figure}[htbp]
\centering
\includegraphics[width=0.325\textwidth]{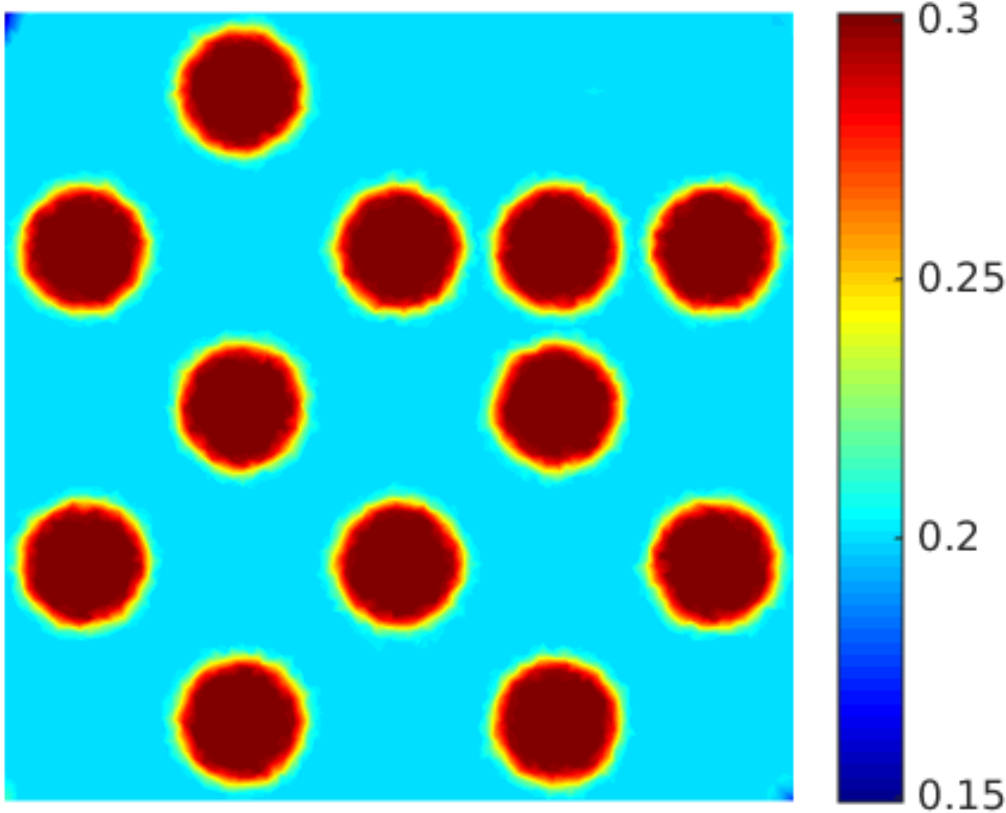}
\includegraphics[width=0.325\textwidth]{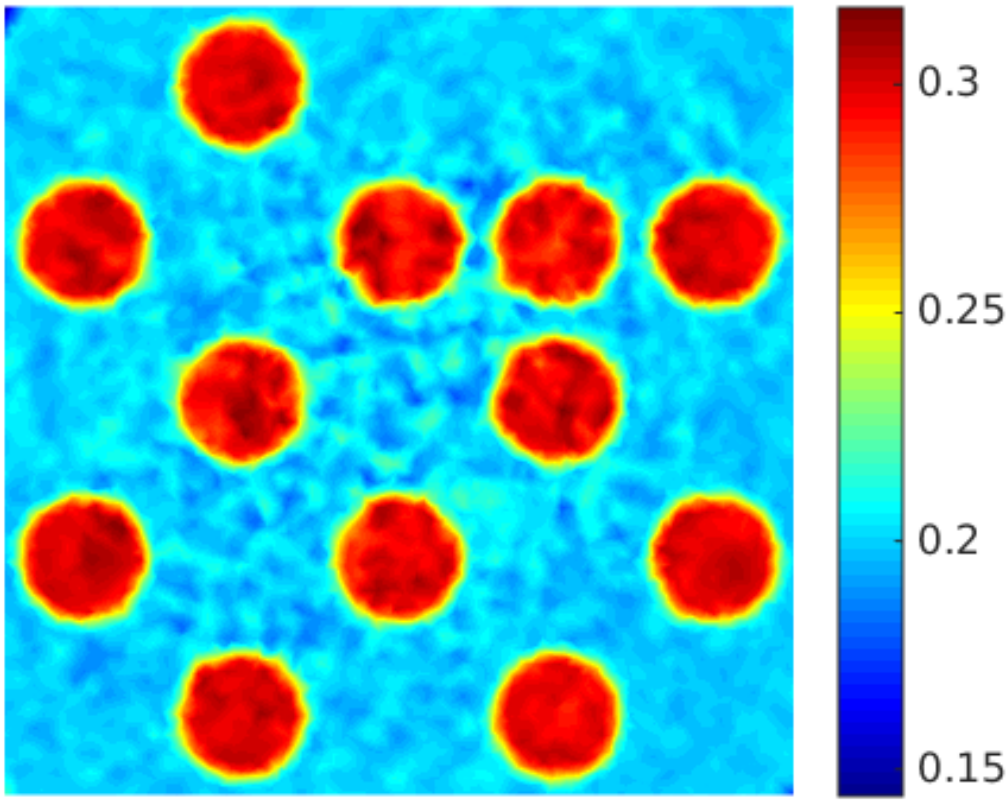}
\includegraphics[width=0.325\textwidth]{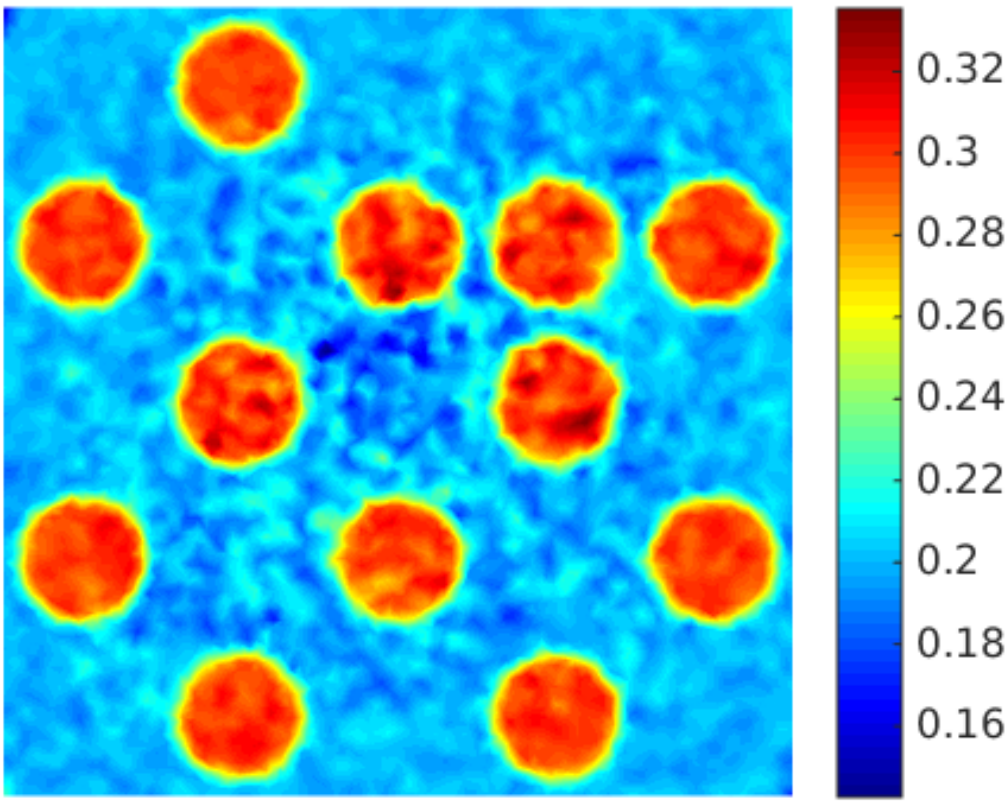}\\
\includegraphics[width=0.325\textwidth]{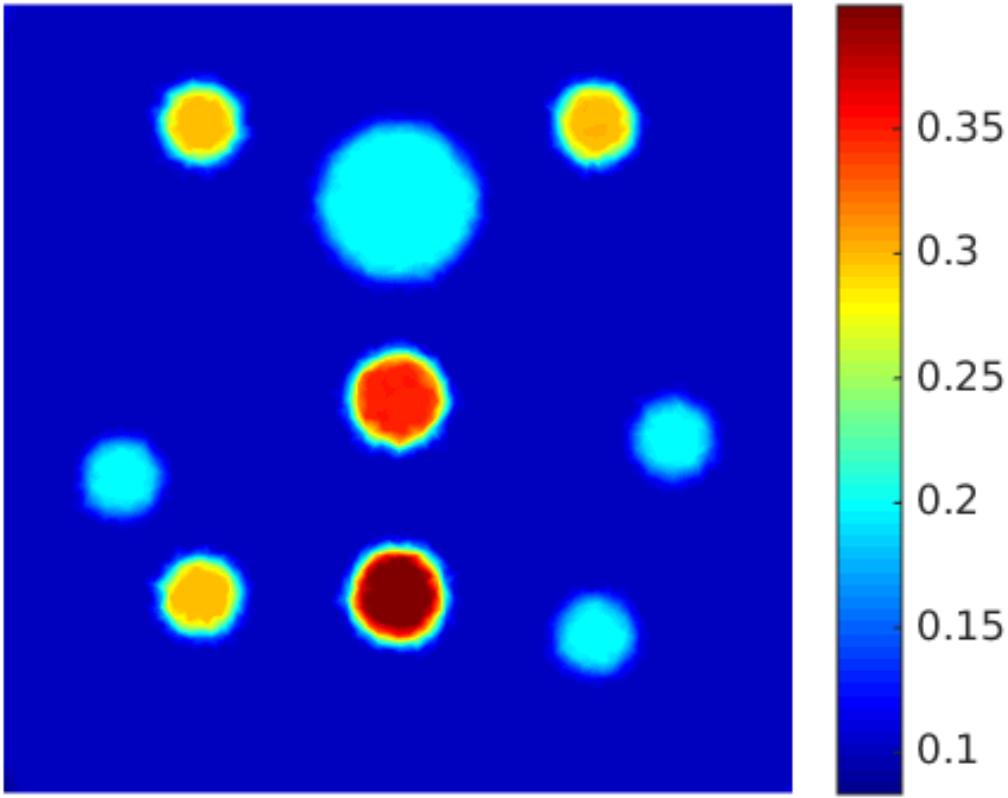}
\includegraphics[width=0.325\textwidth]{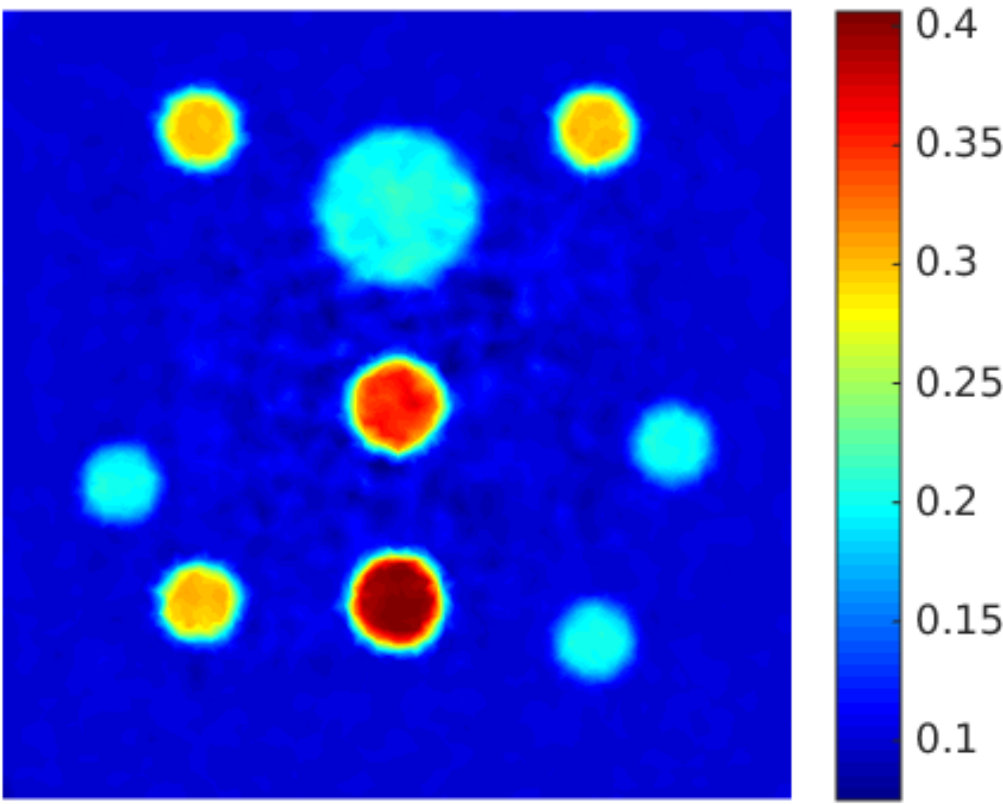}
\includegraphics[width=0.325\textwidth]{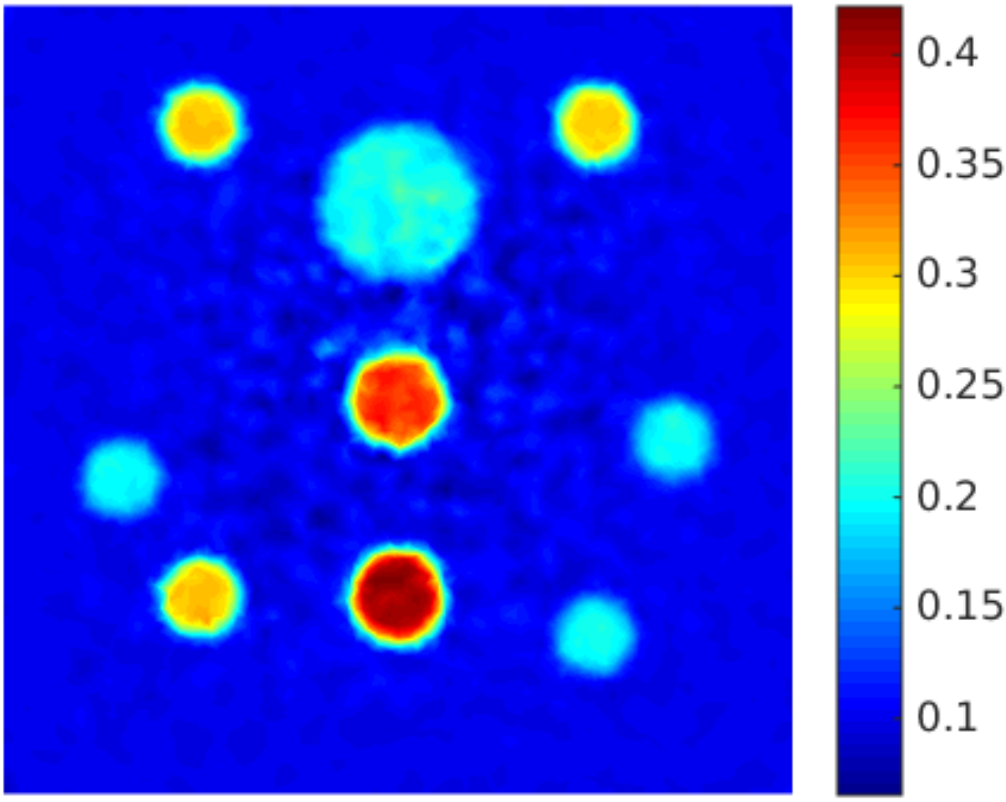}
\caption{Reconstruction of the absorption coefficient pair $\sigma$ (top row) and $\mu\times U_0$ (bottom row) in Figure~\ref{FIG:True Coeff 2Step2} using data at different noise levels (NSR $\eta=0.00$, $\eta=0.05$ and $\eta=0.10$ from left to right).}
\label{FIG:Sigma Mu Direct 2Step2}
\end{figure}

To separate $\mu$ from $\sigma$ in the initial pressure fields, we perform quantitative reconstructions using Algorithm~\ref{ALG:QPAT}. In Figure~\ref{FIG:Sigma Mu Direct 2Step2}, we show the reconstructions from data with SNR $\eta=0.00$ (left), $\eta=0.05$ (middle) and $\eta=0.10$ (right). The relative $L^2$ errors in the reconstructions are $(2.19, 5.26)\times 10^{-2}$, $(4.31, 7.73)\times 10^{-2}$, and $(5.60, 9.34)\times 10^{-2}$ respectively. Once again, we see good separation of the two different absorption coefficients which were mixed together in the initial pressure fields $H$ in Figure~\ref{FIG:H Recon Illu5}. The quantitative reconstruction results are summarized in Table~\ref{TAB:Exp II}, fourth column.

\begin{table}[!ht]
\centering
\setlength{\tabcolsep}{1em}
\renewcommand\arraystretch{1}
\begin{tabular}{c|c|c|c}
NSR & Illumination & $\cE_{L^2}(H)$ & $(\cE_{L^2}(\sigma),\cE_{L^2}(\mu))$\\
\hline\hline
\multirow{4}{*}{$\eta=0.00$} & $g_1$ & $9.30\times 10^{-4}$ & \multirow{4}{*}{$(2.19, 5.26)\times 10^{-2}$}\\
			     & $g_2$ & $1.16\times 10^{-3}$ &\\
			     & $g_3$ & $1.17\times 10^{-3}$ &\\
			     & $g_4$ & $9.38\times 10^{-4}$ &\\ \hline
\multirow{4}{*}{$\eta=0.01$} & $g_1$ & $8.09\times 10^{-3}$ & \multirow{4}{*}{$(2.77, 5.78)\times 10^{-2}$}\\
			     & $g_2$ & $8.25\times 10^{-3}$ &\\
			     & $g_3$ & $7.79\times 10^{-3}$ &\\
			     & $g_4$ & $7.68\times 10^{-3}$ &\\ \hline
\multirow{4}{*}{$\eta=0.05$} & $g_1$ & $1.72\times 10^{-2}$ & \multirow{4}{*}{$(4.31, 7.73)\times 10^{-2}$}\\
			     & $g_2$ & $1.82\times 10^{-2}$ &\\
			     & $g_3$ & $1.82\times 10^{-2}$ &\\
			     & $g_4$ & $1.78\times 10^{-2}$ &\\ \hline
\multirow{4}{*}{$\eta=0.10$} & $g_1$ & $2.51\times 10^{-2}$ & \multirow{4}{*}{$(5.60, 9.34)\times 10^{-2}$}\\
			     & $g_2$ & $2.48\times 10^{-2}$ &\\
			     & $g_3$ & $2.40\times 10^{-2}$ &\\
			     & $g_4$ & $2.79\times 10^{-2}$ &\\ \hline
\end{tabular}
\caption{Quality of reconstructions in Experiment II. Shown are relative $L^2$ errors in the reconstructions of various initial pressure fields (third column) and the corresponding absorption coefficients in Figure~\ref{FIG:True Coeff 2Step2} (fourth column) from ultrasound data with different noise levels.}
\label{TAB:Exp II}
\end{table}

\section{Concluding remarks}
\label{SEC:Concl}

We studied in this paper quantitative image reconstructions in two-photon photoacoustic tomography, aiming at reconstructing the single-photon absorption and the two-photon absorption coefficients of biological tissues from measured ultrasound signals generated by the photoacoustic effect of light absorption. We introduced a nonlinear diffusion equation as the model for light propagation in TP-PAT, and presented a two-step image reconstruction strategy, including a non-iterative quantitative reconstruction step, based on this model. We showed, with computational simulations, that while single-photon absorption and two-photon absorption are mixed in the images of the initial pressure fields, they can be stably separated from each other through the quantitative reconstruction step, using Algorithm~\ref{ALG:QPAT}, even when the ultrasound data contain relatively high level of random noises. Our numerical simulations confirm the results of mathematical analysis of the problem in a previous publication~\cite{ReZh-SIAM17}.

Compared to the case in the regular PAT, quantitative image reconstruction in TP-PAT is far less investigated, theoretically or computationally, to date. Our numerical simulations show great promise in the quantitative imaging of the two-photon absorption. However, there are still a lot of issues that need to be addressed. For instance, it would be very interesting to test the two-step reconstruction method we proposed against experimentally measured data to see what types of resolution and contrast we can get for the two-photon absorption coefficient. It would also be important to develop efficient algorithms to reconstruct the diffusion coefficient $\gamma$ in addition to the absorption coefficients. Last but not the least, reconstructing the Gr\"uneisen coefficient with multispectral data, following for instance the ideas in~\cite{BaRe-IP12,CoArBe-JOSA09,CoLaBe-SPIE09,LaCoZhBe-AO10,RaDiViMaPeKoNt-NP09,RaViNt-OL07,ShCoZe-AO11,YuJi-OL09}, could also be extremely useful as well.

\section*{Acknowledgments}

We would like to thank Professor John C. Schotland (University of Michigan) for useful discussions on the two-photon absorption photoacoustic imaging. This work is partially supported by the National Science Foundation through grants DMS-1321018 and DMS-1620473.


\end{document}